\documentclass[12pt]{article}
\usepackage{epsfig}
\usepackage{graphicx}
\usepackage{a4}
\usepackage{latexsym}
\usepackage{cite}

\usepackage{color}

\usepackage{pslatex}
\usepackage[latin1]{inputenc}
\usepackage[T1]{fontenc}

\renewcommand{\theequation}{\thesection.\arabic{equation}}

\newcommand{\gsim}{\raisebox{-0.07cm}{$\:\:\stackrel{>}{{\scriptstyle
 \sim}}\:\: $} }
\newcommand{\lsim}{\raisebox{-0.07cm}{$\:\:\stackrel{<}{{\scriptstyle
 \sim}}\:\: $} }
\newcommand{\beq}{\begin{equation}}
\newcommand{\eeq}{\end{equation}}
\newcommand{\bea}{\begin{eqnarray}}
\newcommand{\eea}{\end{eqnarray}}
\newcommand{\nn}{\nonumber}
\newcommand{\MSb}{$\overline{\mbox{MS}}$ }

\newcommand{\eqMel}{\raisebox{-0.07cm}{$\stackrel{\rm M}{=} $} }
\newcommand{\hspn}{{\hspace{-4mm}}}
\newcommand{\hspp}{{\hspace{5mm}}}

\newcommand{\Ntil}{\widetilde{\!N}}
\newcommand{\GE}{\gamma_{\rm e}}

\newcommand{\as}{\alpha_{\rm s}}
\newcommand{\ar}{a_{\rm s}}

\def\z#1{{\zeta_{#1}}}
\def\z#1{{\zeta_{\:\! #1}}}
\def\zss{{\zeta_{2}^{\,2}}}
\def\zst{{\zeta_{2}^{\,3}}}
\def\zts{{\zeta_{3}^{\,2}}}

\def\zt{{(1\!-\!z)}}

\def\muR {{\mu_R^{}}}
\def\muF {{\mu_F^{}}}
\def\mH{{m_H^{}}}
\def\muRs{{\mu_R^{\,2}}}
\def\muFs{{\mu_F^{\,2}}}
\def\mHs{{m_H^{\,2}}}
\def\MZs{{M_Z^{\,2}}}

\def\ca{{C^{}_A}}
\def\cas{{C^{\,2}_A}}
\def\cat{{C^{\,3}_A}}
\def\caf{{C^{\,4}_A}}
\def\cf{{C^{}_F}}
\def\cfs{{C^{\, 2}_F}}
\def\cft{{C^{\, 3}_F}}
\def\cff{{C^{\, 4}_F}}
\def\nf{{n^{}_{\! f}}}
\def\nfs{{n^{\,2}_{\! f}}}
\def\nft{{n^{\,3}_{\! f}}}

\def\bz{{\beta_0^{}}}
\def\bn#1{{\beta_0^{#1}}}

\def\lnzt#1{{\ln^{\,#1 \!}\zt}}
\def\xiH{{\xi_H^{(3)}}}
\def\xiD{{\xi_{\rm DY}^{(3)}}}
\def\etaH{{\eta_H^{(3)}}}
\def\etaD{{\eta_{\rm DY}^{(3)}}}

\def\Ag4{{A_{g,4}}}
\def\Dg4{{D_{g,4}}}

\def\pqq(#1){p_{qq}(#1)}
\def\pgg(#1){p_{gg}(#1)}
\def\H(#1){{\rm{H}}_{#1}}
\def\Hh(#1,#2){{\rm{H}}_{#1,#2}}
\def\THh(#1,#2){{\widetilde{\rm{H}}}_{#1,#2}}
\def\Hhh(#1,#2,#3){{\rm{H}}_{#1,#2,#3}}
\def\THhh(#1,#2,#3){{\widetilde{\rm{H}}}_{#1,#2,#3}}

\begin{document}
\setlength{\parskip}{0.15cm}
\setlength{\baselineskip}{0.52cm}

\begin{titlepage}
\renewcommand{\thefootnote}{\fnsymbol{footnote}}
\thispagestyle{empty}
\noindent
DESY 14-123 \hfill August 2014 \\
LPN 14-088 \\
LTH 1014
\vspace{1.0cm}

\begin{center}
{\bf \Large 
Approximate N$^3$LO Higgs-boson production\\[2mm]
cross section using physical-kernel constraints}\\
  \vspace{1.25cm}
{\large
D. de Florian$\,$\footnote{deflo@df.uba.ar}$^{a}$,
J. Mazzitelli$\,$\footnote{jmazzi@df.uba.ar}$^{a,b}$,
S. Moch$\,$\footnote{Sven-Olaf.Moch@desy.de}$^{b,c}$
and
A. Vogt$\,$\footnote{Andreas.Vogt@liverpool.ac.uk}$^{d}$
   \\
}
 \vspace{1.25cm}
 {\it
   $^{a}$ Departamento de F\'\i sica, FCEyN, Universidad de Buenos Aires \\
   (1428) Pabell\'on 1, Ciudad Universitaria, Capital Federal, Argentina \\
   \vspace{0.2cm}
   $^{b}$Deutsches Elektronensynchrotron DESY \\
   Platanenallee 6, D--15738 Zeuthen, Germany \\
   \vspace{0.2cm}
   $^{c}$ II. Institut f\"ur Theoretische Physik, Universit\"at Hamburg \\
   Luruper Chaussee 149, D--22761 Hamburg, Germany \\
   \vspace{0.2cm}
   $^{d}$ Department of Mathematical Sciences, University of Liverpool \\
   Liverpool L69 3BX, United Kingdom \\
 }
  \vspace{1.5cm}
  \large {\bf Abstract}
  \vspace{-0.2cm}
\end{center}
The single-logarithmic enhancement of the physical kernel for Higgs production
by gluon-gluon fusion in the heavy top-quark limit is employed to derive
the leading so far unknown contributions, $\ln^{\,5,\,4,\,3\!} \zt$, to the 
N$^3$LO coefficient function in the threshold expansion.
Also using knowledge from Higgs-exchange DIS to estimate the remaining terms 
not vanishing for $z = \mHs/\hat{s} \to 1$, these results are combined with the 
recently completed soft$\,+\,$virtual contributions to provide an uncertainty 
band for the complete N$^3$LO correction.
For the 2008 MSTW parton distributions these N$^3$LO contributions increase 
the cross section at 14 TeV by 
{$(10 \pm 2)\%$} and {$(3 \pm 2.5)\%$} for the
standard choices $\muR=\mH$ and $\muR=\mH/2$ of the renormalization scale.
The remaining uncertainty arising from the hard-scattering cross
sections can be quantified as no more than 5\%, which is smaller than that due 
to the strong coupling and the parton distributions.

\hfill

\end{titlepage}
\setcounter{footnote}{1}
\renewcommand{\thefootnote}{\fnsymbol{footnote}}

%
\section{Introduction}

After the recent discovery of a new boson by the ATLAS and CMS collaborations
\cite{Aad:2012tfa,Chatrchyan:2012ufa} at the Large Hadron Collider (LHC),
precise theoretical predictions are needed in order to determine whether or not
this particle is indeed, as it appears so far \cite{CMS:yva,ATLAS:2013sla}, the
Standard Model (SM) Higgs boson.
In~particular, to study its properties and to be able to distinguish between SM
and Beyond-the-SM scenarios, it is important to provide precision calculations 
of the Higgs production rate. 

The main production mechanism for the SM Higgs boson at the LHC is the 
gluon-gluon fusion process. 
The radiative corrections in Quantum Chromodynamics (QCD) for the corresponding
inclusive cross section have been computed to next-to-next-to-leading order 
(NNLO) in the effective theory~\cite
{Harlander:2002wh,Anastasiou:2002yz,Ravindran:2003um} 
based on the limit of a large top-quark mass, $m_t \gg \mH$, and later 
for $\mH \lsim 2\,m_t $
 in the full theory~\cite
{Harlander:2009mq,Pak:2009dg,Harlander:2009my}.
The large size of the QCD corrections at this and the previous \cite
{Dawson:1990zj,Djouadi:1991tka,Spira:1995rr,Harlander:2005rq}
order, mainly due to large contributions from the $z \!\to\! 1$ limit, where 
$z$ is the ratio of the Higgs mass $\mH$ to the partonic center-of-mass energy 
$\sqrt{\hat{s}}\,$ squared, $z=\mHs/\hat{s}$, together with the still sizeable 
scale uncertainty have motivated systematic theory improvements beyond NNLO.

At the next-to-next-to-next-to-leading order (N$^3$LO), all plus-distribution 
contributions to the partonic cross section in the \MSb scheme, 
$[\zt^{-1}\ln^{\,k}\zt]_+$ with $0 \le k \le 5$, i.e., the leading 
contributions for Higgs boson production at threshold, are known in the large 
top-mass limit~\cite{MV2005}.
Recently also the  corresponding terms proportional to $\delta \zt$ have been 
computed \cite{Anastasiou:2014vaa} which include the 3-loop virtual 
contributions. 
In Mellin $N$-space, with $N$ being the conjugate variable of $z$, the 
threshold logarithms appear as $\ln^{\,k} N$ with $1\le k \le 2n$ at the
$n$-th order, while the virtual contributions lead to a constant in $N$.
Based on comparisons at the previous orders, the soft-virtual~(SV) 
approximation in $N$-space (which can be supplemented by an all-order 
resummation of threshold contributions up to next-to-next-to-next-to-leading 
logarithmic (N$^3$LL) accuracy \cite{MVV7}) has been shown to yield reliable 
predictions for the total Higgs production cross section, see, e.g., Refs.~%
\cite{Catani:2003zt,MV2005,deFlorian:2012yg,deFlorian:2012za,Bonvini:2014joa}. 
Studies in the soft-collinear effective theory (SCET) have reached similar 
conclusions concerning the validity of an approximation based on threshold 
logarithms~\cite{Ahrens:2008nc,Ahrens:2010rs}.

In this paper we present N$^3$LO and N$^4$LO results beyond the SV 
approximation. 
For a scheme-independent description of the hard scattering process one can
employ physical evolution kernels (also called physical anomalous dimensions) 
which arise from standard QCD factorization once the parton densities (PDFs)
are eliminated from the evolution equation for the physical cross section.
Since the physical evolution kernels exhibit only a single-logarithmic 
enhancement at large $z$, see Refs.~\cite{MV2009c,SMVV}, we are able to 
establish constraints on the coefficient functions in the \MSb scheme.
In this manner we obtain at N$^3$LO the subleading logarithmic contributions 
$\ln^{\,k} \zt$ (or in Mellin space $N^{\,-1}\ln^{\,k}N\,$) for $k=5,\,4,\,3$ 
to the gluon-gluon partonic cross section.
In addition, with the help of results for inclusive deep-inelastic scattering 
(DIS) by Higgs exchange which are known to N$^3$LO~\cite{SMVV}, we can also 
systematically estimate the size of the remaining ${\cal O}(N^{\,-1})$ terms.

Based on the SV contributions together with the new subleading double 
logarithmically enhanced $N^{-1}\ln^{\,k}N$ terms, we are then able to provide 
improved predictions for the yet unknown full N$^3$LO corrections to the 
gluon-gluon coefficient function for inclusive Higgs production.
As an additional uncertainty estimate we study the numerical impact of the 
%
N$^4$LO corrections in the SV approximation. Our analytical results at 
N$^3$LO can be compared to previous phenomenologically motivated approximations
for the third-order cross section \cite{Ball:2013bra,Bonvini:2014jma}.

Beyond the $\zt^0$ terms in the expansion about $z=1$, the gluon-gluon
coefficient function receives `flavour-singlet' contributions which, unlike for
DIS and semi-inclusive $e^+e^-$ annihilation (SIA), cannot be analyzed (so far)
in terms of physical kernels for hadron-collider observables.
Hence an extension of the above results to all powers of $\zt$  along the lines
of Ref.~\cite{MV2009c} can be performed only for the `non-singlet'
$C_A^{\,k} \, n_{\! f}^{\,\ell}$ contributions.
Yet the corresponding terms can, at least, provide useful checks of future
Feynman-diagram calculations.
Finally we take the opportunity to update the corresponding results for the
dominant quark-antiquark annihilation contribution to the Drell-Yan (DY)
process to the same accuracy at N$^3$LO and N$^4$LO.

%
\section{Constraints from the physical evolution kernel}

For $\mH \simeq 125 \mbox{ GeV}$ \cite{Aad:2012tfa,Chatrchyan:2012ufa} the
higher-order corrections can be addressed in the large top-mass approximation, 
in which the effective coupling of the Higgs to partons is given by the 
Lagrangian
\beq
\label{eq:Leff}
  {\cal L}_{\rm{eff}} \:\:=\:\: -\frac{1}{4\upsilon} \, C(\muRs) \: 
  H \, G^{\,a}_{\!\mu\nu} G_{\!a}^{\:\mu\nu} \:\: ,
\eeq
where $\upsilon \simeq 246~\rm{GeV}$ is the Higgs vacuum expectation value and
$G^{\,a}_{\!\mu\nu}$ denotes the gluon field strength tensor.
The matching coefficient $C(\muRs)$ is fully known up to N$^3$LO~\cite
{Chetyrkin:1997un,Schroder:2005hy,Chetyrkin:2005ia}.
Standard QCD factorization, here as usual performed in the \MSb scheme, allows 
to express the inclusive hadronic cross section for Higgs boson production at 
a center-of-mass energy $E_{cm}=\sqrt{S}\,$ as 
\bea
\label{eq:had}
  \sigma(S,\mHs) &\!=\!& \tau\: 
  \sum\limits_{a,b} \:\int_0^1 \frac{dx_1^{}}{x_1^{}} \;\frac{dx_2^{}}{x_2^{}} 
  \; f_{a/h_1^{}}(x_1^{},\muFs)\;f_{b/h_2^{}}(x_2^{},\muFs) 
  \int_0^1 \! dz \;\delta \Big(z - \frac{\tau}{x_1^{} x_2^{}}\Big)\, \times\,  
\nn \\[0.5mm] & & 
  \times\:\: \widetilde{\sigma}_0^{} \: c_{ab}(z,\, \as(\muRs),\, 
  \mHs/\muRs,\, \mHs/\muFs) \;\; ,
\eea
where $\tau=\mHs/S$, and $\muF$ and $\muR$ are the mass-factorization and 
renormalization scales, respectively.  
The PDFs of the colliding hadrons are denoted by $f_{a/h}(x,\muFs)$, the 
subscripts $a,b$ indicating the type of massless parton.
The variable $z=\mHs/\hat{s}\,$ is the partonic equivalent of $\tau$, with 
$\hat{s}=x_1^{}x_2^{}S\,$ being the partonic center-of-mass energy squared.
The complete $\as$-expansion of the effective Higgs-gluon vertex is included 
in $\widetilde{\sigma}_0^{}$, viz
\beq
\label{eq:sigma0}
  \widetilde{\sigma}_0^{} \:\: = \:\:
  \frac{\pi\, C(\muRs)^2}{64\, \upsilon^2} \quad \mbox{ with } \quad
  C(\muRs) \:\: = \:\: - \,\frac{\as(\muRs)}{3 \pi} 
  \: \Big\{ 1 \,+\, 11\: \frac{\as(\muRs)}{4 \pi} \: + \: \ldots \Big\} \;\; .
\eeq
 
$\!$%
We expand the coefficient functions $c_{ab}$ in powers of the strong coupling
with $\ar \equiv \as(\muRs)/(4\pi)$,
\beq
\label{eq:cexp}
 c_{ab}(z,\,\as(\muRs),\, \mHs/\muRs,\, \mHs/\muFs) \:\:=\:\: \sum_{n=0}^\infty
 a_{\rm s}^{\,n} c_{ab}^{(n)}(z,\,\mHs/\muRs,\,\mHs/\muFs) \:\: .
\eeq
At leading order (LO) we have 
$c_{ab}^{(0)} = \delta_{ag}\,\delta_{bg}\,\delta \zt$; at $n\!\geq\!1$ the 
coefficient functions $c_{ab}^{(n)}$ in Eq.~(\ref{eq:cexp}) differ from the 
quantities $\Delta_{\,ab}$ in Refs.~\cite {Anastasiou:2002yz,Ravindran:2003um} 
by a factor of $z^{\,-1}$, cf.~Eq.~(4.3) of \cite{Ravindran:2003um}. 
As~mentioned above, the QCD corrections within the large top-mass limit are 
known up to NNLO~\cite{Anastasiou:2002yz,Harlander:2002wh,Ravindran:2003um}, 
while at N$^3$LO only the soft and virtual (SV) contributions, i.e., the 
plus-distributions ${\cal D}_k(z)=[(1\!-\!z)^{-1}\ln^k(1\!-\!z)]_+$ and the 
$\delta(1\!-\!z)$ terms in the gluon-gluon channel are available so far
\cite{MV2005,Anastasiou:2014vaa}.
Very recently, also the leading double-logarithmic threshold contribution to 
the quark-gluon coefficient function $c_{qg}^{\,(3)}$ has been obtained as part
of an all-order result \cite{LPAV2014}.

\pagebreak

More information about large-$z$ contributions to the N$^3$LO coefficient
function $c_{gg}^{(3)}$ and its higher-order counterparts can be extracted 
from the physical evolution kernel.  
To that end, we consider the case $\muF = \muR = \mH$ (the scale-dependent 
terms can be reconstructed by renormalization-group arguments) and define 
dimensionless partonic `structure functions' ${\cal F}_{ab}$
\beq
  \sigma(S,\mHs) \:\: = \:\: \sum_{a,b} \, \widetilde{\sigma}_0^{} \: 
  {\cal F}_{ab} \:\: .
\eeq
For the sub-dominant $\zt^0$ terms we can restrict ourselves to the 
`non-singlet' case where only the coefficient function $c_{gg}$ and the 
splitting function $P_{gg}$ are taken into account; other contributions are
suppressed by two powers of $\zt$ relative to the leading $\zt^{\,-1}$ terms.
Exploiting the evolution equations for $\as$ and the PDFs one arrives at the 
expression, cf.~Ref.~\cite{MV2009c},
\bea
\label{eq:physkern}
  \frac{d}{d\ln \mHs}\:{\cal F}_{gg} &\!=\!&
  \left\{ 2 P_{gg}(a_s) + \beta(a_s) \: \frac{d c_{gg}(a_s)}{d a_s} \,\otimes\,
  \left(c_{gg}(a_s)\right)^{-1} \right\} \otimes {\cal F}_{gg}
\nn \\
 &\!\equiv\!& 
  K_{gg}\otimes{\cal F}_{gg} \;\equiv\; 
  \sum_{\ell=0}^\infty \, a_s^{\,\ell+1} K_{gg}^{\,(\ell)}\otimes {\cal F}_{gg}
\nn\\
 &\,=\,& \left\{ 2 a_s P_{gg}^{\,(0)} + \sum_{\ell=1}^\infty \, a_s^{\,\ell+1}
  \left(
  2 P_{gg}^{\,(\ell)} - \sum_{k=0}^{\ell-1} \,\beta_k\, 
  \tilde{c}_{gg}^{\,(\ell-k)} \right) \right\} \otimes {\cal F}_{gg}
\eea
which defines the physical evolution kernel $K_{gg}$ and its perturbative
expansion.
Here $\otimes$ denotes the usual Mellin convolution, cf.~Eq.~(\ref{eq:had}), 
while $\beta(a_s)$ stands for the standard QCD beta function, $\beta(\ar) = 
- \beta_{0\,}^{} \ar^{\,2} - \ldots$ with $\beta_0 = 11/3\;C_A - 2/3\;\nf\,$.
$P_{gg}^{\,(\ell)}$ are the $(\ell+1)$-loop gluon-gluon splitting functions, 
defined analogously to $K_{gg}^{\,(\ell)}$ in the middle line of 
Eq.~(\ref{eq:physkern}).
Up to N$^4$LO the expansion coefficients $\tilde{c}_{gg}^{\,(\ell)}$ in the 
last line are given by \cite{vanNeerven:2001pe}
\bea
\label{eq:ctilde}
  \tilde{c}_{gg}^{\,(1)} &\!=\!& c_{gg}^{(1)} 
\; , \nn \\[0.5mm]
  \tilde{c}_{gg}^{\,(2)} &\!=\!& 2c_{gg}^{(2)}-c_{gg}^{(1)}\otimes c_{gg}^{(1)}
\; , \nn \\[0.5mm]
  \tilde{c}_{gg}^{\,(3)} &\!=\!& 
   3c_{gg}^{(3)}
  -3c_{gg}^{(2)}\otimes c_{gg}^{(1)}
  + c_{gg}^{(1)}\otimes c_{gg}^{(1)}\otimes c_{gg}^{(1)}
\; , \nn \\[0.5mm]
  \tilde{c}_{gg}^{\,(4)} &\!=\!&
   4c_{gg}^{(4)}
  -4c_{gg}^{(3)}\otimes c_{gg}^{(1)}
  -2c_{gg}^{(2)}\otimes c_{gg}^{(2)}
  +4c_{gg}^{(2)}\otimes c_{gg}^{(1)}\otimes c_{gg}^{(1)}
  - c_{gg}^{(1)}\otimes c_{gg}^{(1)}\otimes c_{gg}^{(1)}\otimes c_{gg}^{(1)}
\; .  \qquad
\eea
 
The calculation of the physical kernel, given the fact that it contains several
convolutions, is best carried out in $N$-space. The Mellin $N$-moments are 
defined as
\beq
\label{eq:Mtrf}
  f(N) \:\:=\:\: \int_0^1 \! dz \,
  \left(\, z^{\,N-1} \{ - 1 \} \right) \: f(z)_{\{+\}} 
  \:\: ,
\eeq
where the parts in curly brackets apply to plus-distributions. 
A useful if approximate dictionary between the logarithms in $z$-space and 
$N$-space is 
\bea
\label{eq:LogTrf}
  (-1)^{k}\, \Bigg( {\ln^{\,k-1\!}\zt \over 1-z} \Bigg)_{\!+} \!\!\!
  &\eqMel& {1 \over k}\, \bigg( [S_{1-}(N)]^{\,k}
     \;+\; {1 \over 2}\, k (k-1) \z2\, [S_{1-}(N)]^{\,k-2}
     \;+\; O ( [S_{1-}(N)]^{\,k-3} ) \!\bigg) \; , 
\nn \\[0.5mm]
   (-1)^{k}\, \ln^{\,k\!}\zt \hspp \!\!
  &\eqMel& {1 \over N}\, \bigg( \ln^{\,k} \Ntil
    \;+\; {1 \over 2}\, k (k-1) \zeta_{\,2}\, \ln^{\,k-2} \Ntil
    \;+\; O ( \ln^{\,k-3} \Ntil ) \!\bigg) 
  \;+\; O \left( {1 \over N^{\,2}} \right) 
\nn \\
\eea
with $S_{1-}(N) = \ln\,\Ntil - 1/(2N) + O(1/N^{\,2})$ and $\:\Ntil \,=\, 
N e^{\,\GE}$, i.e., $\ln\,\Ntil = \ln N + \GE$ with $\GE \simeq 0.577216$.
Here $\eqMel$ indicates that the right-hand-side is the Mellin transform 
(\ref{eq:Mtrf}) of the previous expression.
The splitting functions, coefficients functions and their products in Mellin 
space can be expressed in terms of harmonic sums~\cite{Vermaseren:1998uu}.
These give rise to harmonic polylogarithms~\cite{Remiddi:1999ew} in $z$-space 
from which one can then extract the large-$z$ and large-$N$ expansions. 
All these manipulations were carried out using the symbolic manipulation system
{\sc Form}~\cite{FORM3,TFORM,FORM4}.

The crucial feature of the (factorization scheme independent) physical 
evolution kernels to be exploited here is the fact that they display only a 
single-logarithmic large-$z$ enhancement. 
This behaviour is in striking contrast to that of the \MSb scheme coefficient
functions, which do include double-logarithmic contributions, i.e., 
$\ln^{\,k\!}\zt$ with $k > n \ge 1$ at N$^n$LO, at all orders in the expansion 
around $z =1$.
This behaviour of the physical evolution kernels has been observed at higher 
orders in perturbative QCD for a variety of observables in DIS, semi-inclusive 
$e^+e^-$ annihilation (SIA) and DY lepton-pair production~\cite{MV2009c,SMVV}. 
For DIS and SIA it can be derived from properties of the unfactorized partonic 
cross sections in dimensional regularization, see Refs.~\cite{AV2010,ASV}.

Also the kernel $K_{gg}$ in Eq.~(\ref{eq:physkern}) is single-log enhanced as
far as it is known so far, i.e., to NNLO. It is therefore plausible to 
conjecture this behaviour to all orders in $\as$.
In particular, requiring the cancellation of the $\lnzt5$ and $\lnzt4$ terms 
in the third line of Eq.~(\ref{eq:ctilde}), we can determine the corresponding 
coefficients of $c_{gg}^{(3)}$. 
Moreover, we observe that the leading large-$N\,$ logarithms of $K_{gg}$ take 
a simple form for the sub-dominant $N^{\,-1}$ contributions,
\bea
\label{eq:xiH}
  \left. K_{gg}^{(1)}\right|_{N^{\,-1}} &\!=\!& 
  - \,\left( 8\,\beta_0\, \ca + 32\,\cas \right)\, \ln N \:+\; {\cal O}(1)
\:\: , 
\nn \\[1mm]
  \left. K_{gg}^{(2)}\right|_{N^{\,-1}} &\!=\!&
  - \,\left( 16\,\beta_0^2\, \ca + 112\,\beta_0\, \cas \right)\, \ln^{\,2\!} N 
  \:+\; {\cal O}( \ln N)
\:\: , 
\nn \\[1mm]
  \left. K_{gg}^{(3)}\right|_{N^{\,-1}} &\!=\!&
  - \,\left( 32\,\beta_0^3\, \ca + \xiH\, \beta_0^2\, \cas \right)\,
  \ln^{\,3\!} N \:+\; {\cal O}( \ln^{\,2\!} N)
\,,
\eea
where the first two lines follow from the NLO and NNLO coefficient functions 
known from the respective diagram calculations in Refs.~\cite
{Dawson:1990zj,Djouadi:1991tka} and~\cite
{Anastasiou:2002yz,Harlander:2002wh,Ravindran:2003um}.
The last line is an obvious generalization based on the results for DIS 
(where the leading-$\beta_0$ coefficients can be derived from the large-$\nf$ 
results in Ref.~\cite{Mankiewicz:1997gz} to all orders) 
and DY, where the coefficients are the same except for $C_A \!\to\! C_F$, see 
%
Eq.~(6.17) of Ref.~\cite {MV2009c}. The unknown coefficient $\xiH$ can be 
estimated by comparing Eq.~(\ref{eq:xiH}) and its completely known analogue in 
DIS, given by Eq.~(5.2) of Ref.~\cite{MV2009c}, together with the Pad\'e 
approximants for the N$^3$LO terms in both equations as about 300 with a 
conservative uncertainty of 50\%, i.e., 150.
This result provides the information about the $\lnzt3$ term of the N$^3$LO
coefficient function. Note that the splitting functions in 
Eq.~(\ref{eq:physkern}) do not contribute to Eq.~(\ref{eq:xiH}) beyond NLO, as 
the diagonal quantities and $P_{qq}^{\,(n)}$ and $P_{gg}^{\,(n)}$ do not show 
any logarithmic higher-order enhancement of the $N^{\,0}$ and $N^{\,-1}$ terms
\cite{Korchemsky:1989si,MVV3,MVV4,DMS05}.

Eqs.~(\ref{eq:physkern}) -- (\ref{eq:xiH}) with 
$\left. K_{gg}^{(3)}\right|_{N^{\,-1}} = {\cal O} (\ln^{\,4\!}N)$ lead to the 
N$^3$LO and N$^4$LO predictions 
\bea
\label{eq:c3z}
  c_{gg}^{(3)}(z) &\!=\!& 
  c_{gg}^{(3)}(z)\Big|_{{\cal D}_k,\delta\zt}
  \:-\: 512 \, \* \cat\: \* \lnzt5
  \:+\: \Big\{
    1728\, \* \cat + \frac{640}{3}\: \* \cas\, \* \bz
  \Big\} \, \* \lnzt4
\nn \\[1mm]
 & & \mbox{\hspn} 
   +\: \left\{
     \left( - \,\frac{1168}{3} + 3584\, \* \z2 \right) \* \cat
   - \left( \frac{2512}{3} + \frac{1}{3}\: \* \xiH \right) \* \cas\, \* \bz
   - \frac{64}{3}\: \* \ca\, \* \bn2
  \right\}\, \* \lnzt3
\nn \\[1mm]
 & & \mbox{\hspn} +\: {\cal O}\Big( \ln^{\,2\!}\zt \Big)
\eea
and
\bea
\label{eq:c4z}
  c_{gg}^{(4)}(z) &\!=\!&
  c_{gg}^{(4)}(z)\Big|_{{\cal D}_k,\delta\zt}
  \:-\: \frac{4096}{3}\: \* \caf\: \* \lnzt7
  + \left\{
    \frac{19712}{3}\: \* \caf + \frac{3584}{3}\: \* \cat\, \* \bz
  \right\}\, \* \lnzt6
\nn \\[1mm] && \mbox{\hspn} 
  + \:\left\{
      \left( -\, 2240 + 23552\, \* \z2 \right) \* \caf
    - \left(\frac{19136}{3} + \frac{8}{3}\: \* \xiH \right) \* \cat\,\* \bz
    - \frac{1024}{3}\: \* \cas\, \* \bn2
  \right\}\, \* \lnzt5
\nn \\[1mm]
 & & \mbox{\hspn} +\: {\cal O}\Big( \ln^{\,4\!}\zt \Big)
\eea
at $\,\muR \,=\, \muF \,=\, \mH$,
where $c_{gg}^{(n)}(z)\big|_{{\cal D}_k,\delta \zt}$ denotes the $z$-space SV 
approximation at N$^n$LO. The coefficients for $n=3$ can be found in 
Eqs.~(17) -- (22) of Ref.~\cite{MV2005} and Eq.~(10) of Ref.~\cite
{Anastasiou:2014vaa} (where the expansion is in powers of $\as/\pi$ instead of
our $\ar = \as/(4\pi)$).
The coefficients multiplying leading and next-to-leading $\ln^{\,k\!}\zt$ terms 
in Eq.~(\ref{eq:c3z}) and (\ref{eq:c4z}) agree with those for DY case in 
Eqs.~(6.24) and (6.25) in Ref.~\cite{MV2009c} if $C_F$ is replaced by $C_A$
in the latter results.  For the third logarithm this is, unsurprisingly, only 
true for the $\beta_0^2$ contribution. 
The leading $\ln^{\,k\!}\zt$ terms in Eq.~(\ref{eq:c3z}) and (\ref{eq:c4z}) 
agree with the old conjecture of Ref.~\cite{Kramer:1996iq}, i.e., the 
coefficients of $\ln^{\,2n-1\!}\zt$ and ${\cal D}_{\,2n-1}$ are the same at 
N$^n$LO up to a sign. 
On the other hand, the subleading terms in Eq.~(\ref{eq:c3z}) do not agree with
the phenomenological ansatz employed in Refs.~\cite
{Ball:2013bra,Bonvini:2014jma}.

Seven of the eight plus-distributions of the N$^4$LO SV contribution 
$c_{gg}^{(4)}(z)\big|_{{\cal D}_k,\delta \zt}$ in Eq.~(\ref{eq:c4z}) can be 
obtained by expanding and Mellin inverting the result of the N$^3$LO + N$^3$LL 
soft-gluon exponentiation. 
The coefficients of ${\cal D}_k$ for $2 \leq k \leq 7$ can be found in Eq.~(16)
of Ref.~\cite{Ravindran:2006cg} and that of ${\cal D}_{\,1}$ in Eq.~(13) of 
Ref.~\cite{Ahmed:2014cla}. 
The remaining ${\cal D}_{\,0}$ and $\delta \zt$ terms, on the other hand, 
require a fourth-order calculation. The ${\cal D}_{\,0}$ term can be predicted 
up to two unknown anomalous dimensions at four loops which are usually denoted 
by $A_{g,4}$ and $D_{g,4}$, see, e.g., Refs.~\cite{MV2005,MVV7}, as
\bea
\label{eq:d0at4loops}
\lefteqn{ c_{gg}^{(4)}\Big|_{{\cal D}_0} \; = \; 
  \Dg4
  \:+\: \caf \* \Bigg(
    - \frac{50096}{9}
    + \frac{11328416}{729}\, \* \z2
    + \frac{8392600}{81}\, \* \z3
    + \frac{1581760}{81}\, \* \zss
    + \frac{3461120}{9}\, \* \z5
} \nn \\[0.5mm] & & \mbox{\hspp}
    - \frac{6894080}{27}\, \* \z2 \* \z3
    + \frac{372416}{15}\, \* \zst
    - 217184\, \* \zts
    - \frac{595616}{15}\, \* \zss \* \z3
    - 562176\, \* \z2 \* \z5
    + 983040\, \* \z7
  \Bigg)
\nn\\[1mm] & & \mbox{}
  \:+\: \cat\, \* \nf \* \Bigg(
      \frac{191776}{81}
    - \frac{3613696}{729}\, \* \z2
    - \frac{2285696}{81}\, \* \z3
    - \frac{401920}{81}\, \* \zss
    + \frac{492800}{9}\, \* \z2 \* \z3
\nn\\[0.5mm] & & \mbox{\hspp}
    - \frac{729088}{9}\, \* \z5
    -\frac{69248}{15}\, \* \zst
    + 30400\, \* \zts
  \Bigg)
\nn\\[0.5mm] & & \mbox{}
  \:+\: \cas\, \* \nfs \* \Bigg(
    - \frac{17920}{81}
    + \frac{290816}{729}\, \* \z2
    + \frac{89344}{81}\, \* \z3
    + \frac{2560}{9}\, \* \zss
    - \frac{69376}{27}\, \* \z2 \* \z3
    + \frac{32768}{9}\, \* \z5
  \Bigg)
\nn\\[0.5mm] & & \mbox{}
  \:+\: \cas\, \* \cf\, \* \nf \* \Bigg(
      \frac{108272}{81}
    - \frac{62752}{27}\, \* \z2
    - \frac{340712}{27}\, \* \z3
    - 256\, \* \zss
    + \frac{13312}{9}\, \* \z2 \* \z3
    + \frac{512}{5}\, \* \zst
    + 9088\, \* \zts
  \Bigg)
\nn\\[1mm] & & \mbox{}
  \:+\: \ca\, \* \cf\, \* \nfs \* \Bigg(
    - \frac{15008}{81}
    + \frac{2144}{9}\, \* \z2
    + \frac{3584}{27}\, \* \z3
    - \frac{512}{3}\, \* \z2 \* \z3
  \Bigg)
\:\: .
\eea
The derivation of the this result required the extension of the calculations of
Ref.~\cite {MV2005} to the $\as^{\,4}$ part of the exponentiation function
$g_5^{}$, see also Refs.~\cite{AV2001,Catani:2003zt}.

The coefficient $A_{g,4}$ has been estimated by Pad\'e approximants as
$A_{g,4} \,=\, ( 17.7,\: 9.70,\: 3.49)\cdot 10^{\,3}$ for 
$\nf \,=\, 3,\: 4,\: 5$ effectively massless flavours. A corresponding estimate
for $D_{g,4}$ is
\beq
\label{eq:d4pade}
  D_{g,4}(\nf=3) \;=\;  12  \cdot 10^{\,5}\: , \quad
  D_{g,4}(\nf=4) \;=\;  9.3 \cdot 10^{\,5}\: , \quad
  D_{g,4}(\nf=5) \;=\;  6.8 \cdot 10^{\,5}\: ,
\eeq
which is less reliable, as due to $D_{g,1} = 0$ only the two coefficients of
Refs.~\cite{AV2001,Catani:2003zt,MV2005,Laenen:2005uz,Idilbi:2005ni}
are available.
Corresponding estimates for the quark quantities $A_{q,4}$ and $D_{q,4}$ 
relevant to the Drell-Yan process can be obtained by multiplying the above 
results by $C_F/C_A$.  

Using Eqs.~(\ref{eq:LogTrf}), our new result (\ref{eq:c3z}) together with the
coefficients of $c_{gg}^{(3)}(z)\big|_{{\cal D}_k,\delta \zt}$ in Refs.~\cite
{MV2005,Anastasiou:2014vaa} can be employed to rigorously extend the $N$-space 
N$^3$LO threshold expansion to 
\bea
\label{eq:c3gN}
  \kappa_3^{}\, c_{gg}^{(3)}(N) \!&\!\simeq\!&
    1.152\, \ln^{\,6\!}N
  + 5.46171\, \ln^{\,5\!}N
  + 23.8352\, \ln^{\,4\!}N
  + 44.9659\, \ln^{\,3\!}N
\\[1mm] & & \mbox{\hspn}
  + 85.6361\, \ln^{\,2\!}N
  + 60.7085\, \ln N
  + 57.0781
\nn \\[0.5mm] & & \mbox{\hspn\hspn\hspn}
  + \: N^{\,-1} \!\left\{
    3.456\,\ln^{\,5\!}N
  + 19.7023\, \ln^{\,4\!}N
  + ( 61.7304 + .0115\,\xi_H^{(3)\,} ) \ln^{\,3\!}N
  + {\cal O} ( \ln^{\,2\!}N )
  \right\}
\nn 
\eea
with $\kappa_3^{} = 1/2000 \simeq 1/(4\pi)^3$.
Here we have inserted the QCD values of the group factors, $C_A = 3$ and 
$C_F = 4/3$, used the physical value of $\nf = 5$ light flavours at scales of 
order $\mHs$, and truncated coefficients including the Riemann $\zeta$-function
and the Euler-Mascheroni constant~$\GE$. 
The factor $\kappa_3^{}$, as $\kappa_4^{}$ in Eq.~(\ref{eq:c4gN}) below, 
approximately converts the coefficients to an expansion in~$\as$.

Note that the $N^{\,-1}$ coefficients receive contributions from both the
plus-distributions and the $\ln^{\,k\!}\zt$ terms of Eq.~(\ref{eq:c3z}), 
hence the $z$-space and $N$-space SV approximations lead to different 
predictions for cross sections.
It is clear from Eq.~(\ref{eq:c3gN}) that the coefficient $\xi_H^{(3)}$ is not 
a major source of uncertainty; its contribution to the coefficient of 
$N^{\,-1}\,\ln^{\,3\!}N$ is expected to be below 10\%.

The N$^4$LO result corresponding to Eq.(\ref{eq:c3gN}) reads, with 
$\kappa_4^{} = 1/25000 \simeq 1/(4\pi)^4$,
\bea
\label{eq:c4gN}
  \kappa_4^{}\, c_{gg}^{(4)}(N) \!&\!\simeq\!&
    0.55296\, \ln^{\,8\!}N
  + 3.96654\, \ln^{\,7\!}N
  + 21.2587\, \ln^{\,6\!}N
  + 62.2985\, \ln^{\,5\!}N
\nn \\[1mm] & & \mbox{\hspn}
  + 150.141\, \ln^{\,4\!}N
  + 212.443\, \ln^{\,3\!}N
  + ( 256.373 + 2\,\kappa_4^{}\,A_{g,4} )\, \ln^{\,2\!}N
\\[1mm] & & \mbox{\hspn}
  + ( 142.548 + \kappa_4^{} \left[ 4\,\GE\, A_{g,4}- D_{g,4} \right] ) \ln N
  \,+\, \kappa_4^{}\, g_{_0,4}
\nn \\ & & \mbox{\hspn\hspn\hspn}
  + \: N^{\,-1} \!\left\{
    2.21184\,\ln^{\,7\!}N
  + 19.6890\,\ln^{\,6\!}N
  + ( 86.4493 + 552\,\kappa_4^{}\,\xi_H^{(3)\,} ) \ln^{\,5\!}N
  + {\cal O} ( \ln^{\,4\!}N )
  \!\right\}
\nn \: .
\eea
Here the coefficient $A_{g,4}$ is practically negligible, its contribution to
the $\ln^{\,2\!}N$ and $\ln N$ coefficients being of the order of 0.1\%.
The uncertainty of $D_{g,4}$ in Eq.~(\ref{eq:d4pade}), conservatively set to 
100\%, is an effect of order $\pm\,$20\% for the $\ln N$ term. 
The constant-$N$ contribution $g_{_0,4}$, i.e., the fourth-order term of the 
prefactor of the soft-gluon exponential, see, e.g., Refs.~\cite{MVV7,AV2001} 
can be estimated by three Pad\'e approximants which yield a fairly wide spread 
of values suggesting $\kappa_4^{}\,g_{_0,4} = 65\pm 65$. 
%
Alternatively this quantity can be estimated via a calculation in which the 
constant-$N$ contributions in the integrals for the soft-gluon exponent,
which we evaluate in the form given by (2.3) -- (2.6) and (3.2) of 
Ref.~\cite{MVV7}, are not discarded. This modified way to write the resummation
formulae leads to much smaller coefficients of the constant-$N$ prefactors 
of the soft-gluon exponential at NNLO and N$^3$LO which can be used to obtain
a range for $g_{_0,4}$ consistent which the one given~above.

Exact SU(N) expressions corresponding to Eq.~(\ref{eq:c3gN}) and the $\ln N$
enhanced parts of Eq.~(\ref{eq:c4gN}) can be found in the Appendix, together
with third- and fourth-order predictions for the respective highest-three
logarithms beyond the $\zt^0$ terms given in Eqs.~(\ref{eq:c3z}) and 
(\ref{eq:c4z}) above.

%
\section{Approximate N$^3$LO phenomenology}
\setcounter{equation}{0}

\vspace{-1.5mm}
Before we address the numerical impact of $N^{\,-1}$ contributions to the 
coefficient function, we briefly discuss the soft$\,+\,$virtual (SV) 
approximation.  
In $z$-space this approximation can be defined by keeping only the 
${\cal D}_k(z)$ and $\delta \zt$ terms in the cross section, cf.\ 
Eq.~(\ref{eq:c3z}). The soft coefficients in $z$-space are affected, however, 
by the artificial presence of factorially-growing subleading terms, originating
in the mis-treatment of kinematic constraints such as energy conservation, 
that spoil the accuracy of the approximation for higher-order predictions at 
limited logarithmic depth \cite{Catani:1996yz}.

\vspace{-0.6mm}
The natural choice for the soft-gluon enhanced contributions is Mellin 
$N$-space, where instead of plus-distributions in $z$ the dominant threshold 
contributions are given by powers of $\ln N$, and the kinematic constraints are
automatically imposed. Consequently the $N$-space SV approximation is defined 
by keeping the terms in the coefficient function that do not vanish for
$N \!\to\! \infty$, cf.~Eq.~(\ref{eq:c3gN}).

\vspace{-0.6mm}
The numerical contributions of the $\ln^{\,k}N$ terms, $0 \leq k \leq 2\:\!n$,
of the Mellin-transformed coefficient functions $c_{gg}^{\,(n)}$ in 
Eq.~(\ref{eq:cexp}) to the cross section (\ref{eq:had}) are illustrated up to 
N$^{n=3}$LO in Table~\ref{table}, where all numbers are normalized to the 
lowest-order result proportional to $[f_{g/p} \otimes f_{g/p}](\tau)$ with 
$\tau = \mHs/S$. 
All these results have been calculated in the heavy-top limits for 
$\mH \,=\, 125 \mbox{ GeV}$, $E_{\rm cm} \,=\, \sqrt{S} \,=\, 14 \mbox{ TeV}$, 
the central gluon distribution $f_{g/p}$ of the 2008 NNLO MSTW set 
%
\cite{Martin:2009iq} and the corresponding value $\as(\MZs) = 0.1171$ of 
the strong coupling leading to $\as(\mHs) = 0.1118$ at $\muF=\muR=\mH$. 
Also shown is the corresponding normalized expansion of the prefactor function 
$[\:\!C(\muRs = \mHs)]^2$ in~Eq.~(\ref{eq:sigma0}).

\vspace{-0.6mm}
All these contributions are positive, as are the $\ln N$ enhanced terms at
N$^4$LO, see Eq.~(\ref{eq:c4gN}). The same is true for the corresponding
coefficient functions for the Drell-Yan process and semi-inclusive $e^+e^-$
annihilation, cf.~Table 1 and Eq.~(37) of Ref.~\cite{MV2009b}, while for DIS
only the $a_s^{\,n}\ln^{\,k \geq n} N$ contributions are positive at $n\leq 4$,
see Table 1 of Ref.~\cite{MVV7}. In all these cases the complete SV result is
smoothly approached when the $\ln^{\,k} N$ terms are included one by one.
This is in contrast to the $z$-space SV approximation which exhibits large
cancellations between the ${\cal D}_k(z)$ contributions as illustrated at
N$^3$LO for DIS in Fig.~4 of Ref.~\cite{MVV7} and for Higgs production in
Ref.~\cite{Anastasiou:2014vaa}.

\begin{table}[tbh]
\begin{center}
\begin{tabular}{|l|c|c|c|c|}
\hline
  & \phantom{XXXXX} & \phantom{XXXXX} 
  & \phantom{XXXXX} & \phantom{XXXXX} \\[-4mm]
  &  LO &  NLO & NNLO & N$^3$LO \\[1mm]
\hline\hline
             &         &          &          &          \\[-4mm]
constant     & $100$   & $77.4$   & $32.2$   & $8.04$   \\[0.5mm]
(delta)      & $(100)$ & $(35.1)$ & $(1.72)$ & $(5.07)$ \\[0.5mm]
$\ln N$      &         & $14.8$   & $12.0$   & $5.14$   \\[0.5mm]
$\ln^2 N$    &         & $7.16$   & $7.56$   & $4.04$   \\[0.5mm]
$\ln^3 N$    &         &          & $1.07$   & $1.09$   \\[0.5mm]
$\ln^4 N$    &         &          & $0.18$   & $0.27$   \\[0.5mm]
$\ln^5 N$    &         &          &          & $0.025$  \\[0.5mm]
$\ln^6 N$    &         &          &          & $0.002$  \\[0.5mm]
\hline
             &         &          &          &          \\[-4mm]
SV           & $100$   & $99.4$   & $53.0$   & $18.6$   \\
\hline\hline
             &         &          &          &          \\[-4mm]
$C^{\,2}(\mHs)$& 100   &  19.6    &  2.05    &  0.12    \\[1mm]
\hline
\end{tabular}
\caption{ \label{table} \small
 The individual contributions of the $\ln^{\,k}N$ terms in the $N$-space 
 coefficient functions $c_{gg}^{(n \leq 3)}$ at $\muR = \muF = \mH$ to the 
 Higgs production cross section for $\mH = 125$ GeV, $E_{\rm cm}= 14$ TeV, and 
 the central gluon density and five-flavour $\as$ of Ref.~\cite{Martin:2009iq}.
 All results are given as percentages of the LO contribution. 
 Also shown, in the same manner, is the expansion of the prefactor function 
 \mbox{$[C(\muRs = \mHs]^2$)}, calculated in the on-shell scheme for the 
 top mass with $m_t^2 = 3.00 \cdot 10^{\,4} \mbox{ GeV}^2$.}
\end{center}
\vspace{-6mm}
\end{table}

\vspace{-0.6mm}
Furthermore the formally leading terms, i.e., those with the highest powers
of $\ln N$, provide numerically small contributions to the cross section;
the dominant part of the threshold corrections arises from the lowest-power
logarithms and the constant terms. This is due to the pattern of coefficients
in, for example, Eq.~(\ref{eq:c3gN}), which is comparable but less pronounced
than that in DIS and SIA, and the low value of $\tau$ for the production of a
125 GeV Higgs-boson at the LHC, which leads a low effective value of $N$ of
$N_{\rm eff} \approx 2$ for the $\ln^{\,k} N$ contributions according to 
Table~\ref{table}.
 
\vspace{-0.6mm}
Another interesting feature shown in Table~1 is the rather large value of the
$\delta \zt$ term at N$^3$LO \cite{Anastasiou:2014vaa} which contributes, for 
the value of $\as$ given above, about three times as much as its NNLO 
counterpart.
It accounts for 63\% of the constant-$N$ contribution at this order, the rest
of which arises from the Mellin transform of the ${\cal D}_k$ terms, such as
the first line of Eq.~(\ref{eq:LogTrf}) for $k = 2$.

\begin{figure}[p]
\begin{center}
  \includegraphics[width=10.8cm]{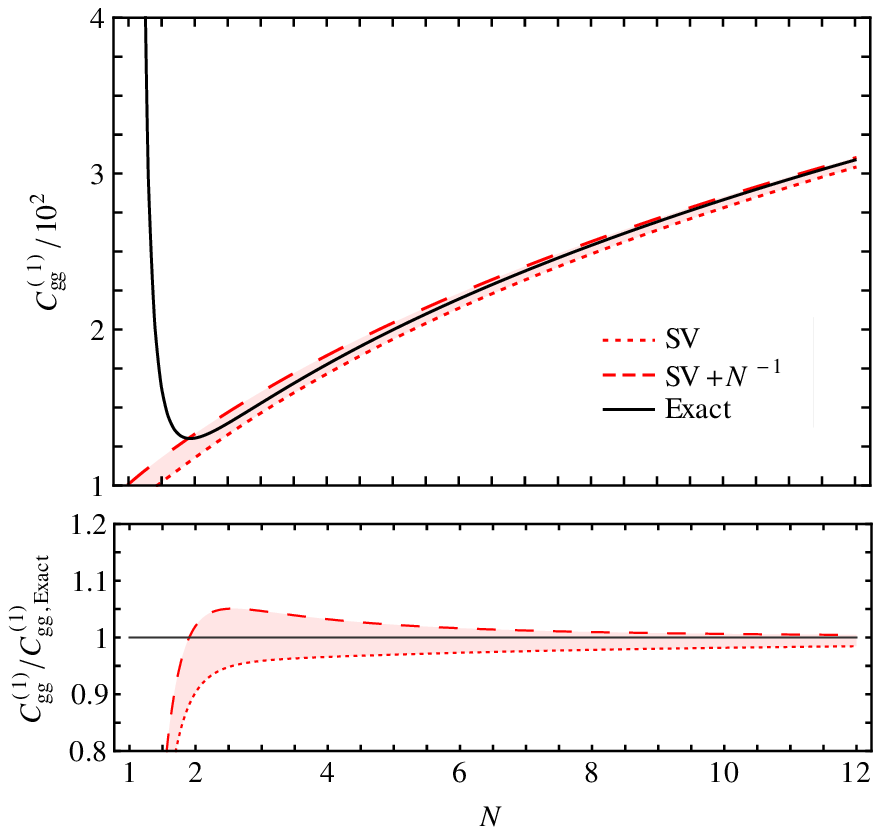}
\\[2mm]
  \includegraphics[width=10.8cm]{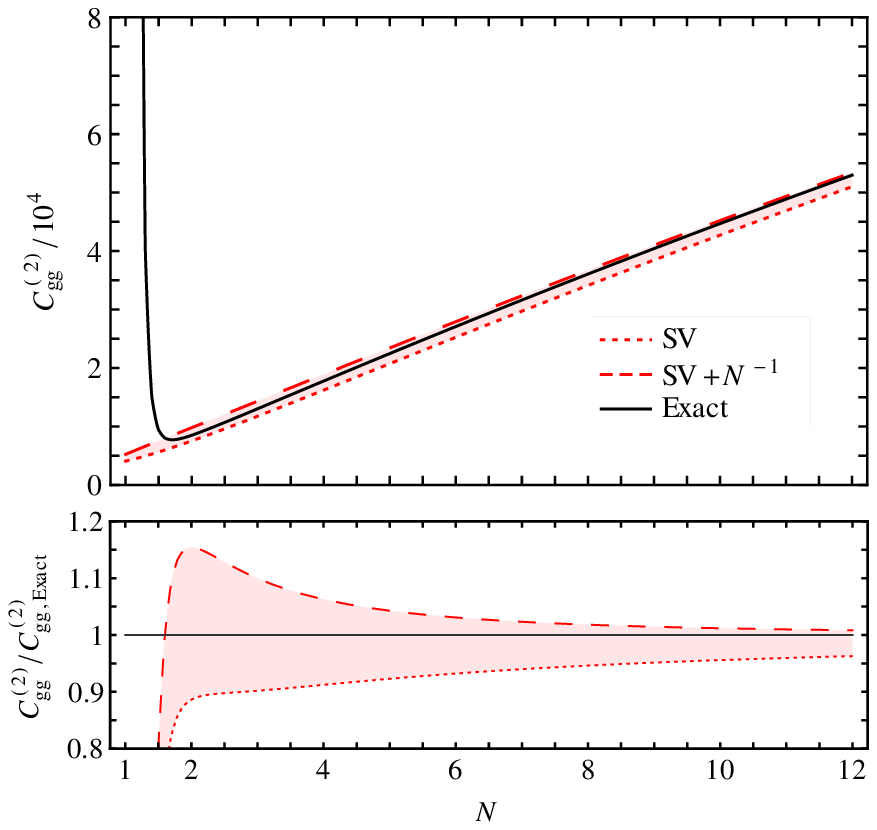}
\end{center}
\vspace{-0.6cm}
\caption{\label{previous} \small
 The exact results for the $N$-space gluon-gluon coefficient functions for 
 $\muR=\muF=\mH$ at NLO (top) and NNLO (bottom) in the heavy-top limit, 
 together with the corresponding SV approximations \mbox{(dotted)} and the 
 SV terms plus the $N^{\,-1}$ contributions (dashed). 
 The respective lower panels show the relative positions and widths of the 
 error bands defined by these two approximations.} 
\vspace{-0.3cm}
\end{figure}

\vspace{-0.6mm}
We are now ready to analyze the effect of adding the subdominant $N^{-1}$ 
contributions to the SV terms.
Before turning to N$^3$LO, we compare the resulting approximation to the exact 
result at NLO and NNLO in Fig.~\ref{previous}. It is clear that including the
$N^{-1}$ terms improves the approximation at large $N$. 
Interestingly, the exact result lies between the SV and the SV$\,+\,N^{-1}$ 
approximations at $N\gsim 2$ at both NLO and NNLO. It is therefore not 
unreasonable to assume that this behaviour also holds at N$^3$LO; hence one 
can constrain $c_{gg}^{(3)}(N)$ even in this region in $N$.

This situation is, in fact, expected from related studies of the DY process 
\cite{MV2009c} and Higgs-exchange DIS~\cite{SMVV}. It is particularly 
interesting to consider the latter case as the coefficient functions are 
completely known to N$^3$LO. Thus,
in order to estimate the size of the $N^{\,-1}$ logarithms not determined in 
Eq.~(\ref{eq:c3gN}), we compare with Ref.~\cite{SMVV} and expand 
the gluon coefficient function $c_{\rm{DIS}}^{(n)}(N)$ of Higgs-exchange DIS 
up to ${\cal O}(N^{-1})$ at both NNLO and N$^3$LO.
We find 
\bea
\label{DIS}
  c_{\rm{DIS}}^{(2)}\Big|_{N^{-1}\ln^{\,k} N}
  &\!\propto\!&
 \ln^3 N + 5.732\, \ln^2 N + 8.244\, \ln N 
  - 3.275
\; , \nn \\[1mm]
 c_{\rm{DIS}}^{(3)}\Big|_{N^{-1}\ln^{\,k} N}
 &\!\propto\!&
 \ln^5 N + 12.65\, \ln^4 N + 52.56\, \ln^3 N + 92.01\, \ln^2 N + 18.13\, \ln N 
  - 24.30 
\qquad
\eea
for $C_A=3$, $C_F=4/3$ and $\nf=5$, where we have normalized the expressions 
such that the coefficient of the leading logarithm is equal to~$1$.
The analogous expressions for Higgs production~are
\bea
\label{HiggsProd}
  c_{gg}^{(2)}\Big|_{N^{-1}\ln^{\,k} N}
  &\propto&
  \ln^3 N + 2.926\, \ln^2 N + 5.970\, \ln N + 2.007
\; , \nn \\[1mm]
  c_{gg}^{(3)}\Big|_{N^{-1}\ln^{\,k} N}
  &\propto&
  \ln^5 N + 5.701\, \ln^4 N 
  + \left( 17.86 + 0.00333\,\xiH \right) \ln^3 N
  + {\cal O}(\ln^2 N)
\; .
\eea
Comparing Eqs.~(\ref{DIS}) and (\ref{HiggsProd}) an interesting pattern emerges:
the size of the coefficients of the non-leading logarithms for Higgs production
is always smaller than that of their analogues for Higgs-exchange DIS; the 
ratio is a factor of about 1/2 or (much) less except for the $\ln^1 N$ terms. 
Thus we suggest as a conservative estimate of the complete $N^{\,-1}$ 
contribution
\beq
\label{eq:c3est}
 c_{gg}^{(3)}\Big|_{N^{-1}\ln^{\,k} N}^{\,\rm estimate}
 \:\propto\:
 \ln^5 N + 5.701\, \ln^4 N + 18.9\, \ln^3 N + 46\, \ln^2 N + 18\,\ln N + 9
\; ,
\eeq
where we have used $\xiH = 300$ as roughly indicated by the physical-kernel
coefficients in Ref.~\cite{MV2009c}.

The above equation includes an estimate of the non-logarithmic $N^{-1}$ 
contribution to $c_{gg}^{(3)}(N)$.
The ratio of the corresponding coefficient to that of $N^{-1} \ln N$ is 
moderate with 0.58 at NLO and 0.34 at NNLO, which may even indicate a trend 
towards lower values if the order is increased. Hence a ratio of 0.5 at 
N$^3$LO, as used in Eq.~(\ref{eq:c3est}), appears to be sufficiently 
conservative 
(recall that these terms contribute positively to the cross section, so for 
larger coefficients we have larger contributions from the estimated terms 
which lead to a wider, i.e., more conservative error band).

\begin{figure}[tb]
\vspace*{-0.3cm}
\begin{center}
  \includegraphics[width=12.4cm]{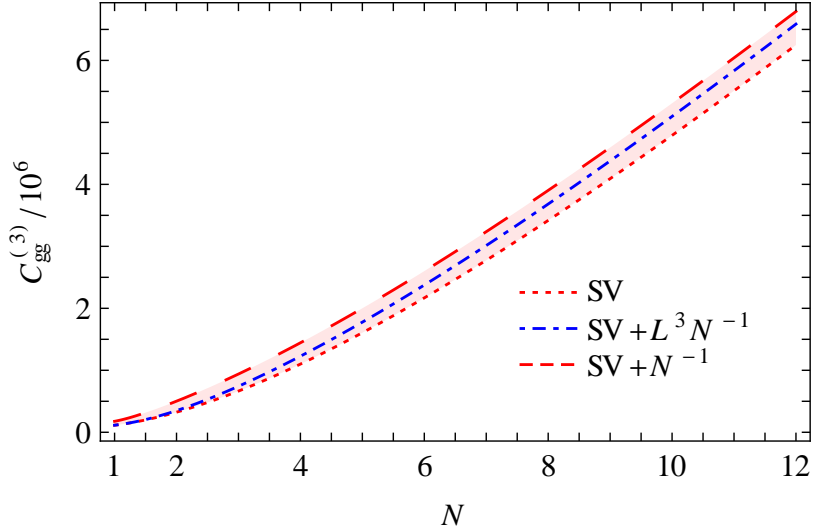}
\end{center}
\vspace{-0.8cm}
\caption{\label{n3lo} \small
 The Mellin-space N$^3$LO coefficient function $c_{gg}^{(3)}(N)$ as 
 approximated, for $N \protect\gsim 2$, by the $N^{\,0}$ SV contributions in 
 Eq.~(\ref{eq:c3gN}) (dotted), 
 the SV contribution plus the three $N^{\,-1} \ln^{\,k\!}N$ terms 
 (approximately) known from physical kernels constraints (dash-dotted), 
 and by the SV terms plus the estimated complete $N^{\,-1}$ contributions in
 Eq.~(\ref{eq:c3est}) (dashed).}
\vspace{-0.3cm}
\end{figure}

Summarizing these constraints, we show in Fig.~\ref{n3lo} the coefficient 
function $c_{gg}^{(3)}(N)$ in the SV approximation, for the SV terms plus the 
$N^{\,-1} \ln^{\,k}N$ contributions with $k \geq 3$ as in Eq.~(\ref{eq:c3gN}), 
and for the SV terms plus the estimate (\ref{eq:c3est}) of all $N^{\,-1}$ 
contributions. Varying the value of $\xiH$ by $\pm 50\%$ has a very small 
impact on the latter two results. 
Based on the pattern observed at NLO and NNLO, we expect that the exact result 
falls in the band displayed in the figure for $N\gsim 2$. 
 
\begin{figure}[p]
\begin{center}
  \includegraphics[width=10.7cm]{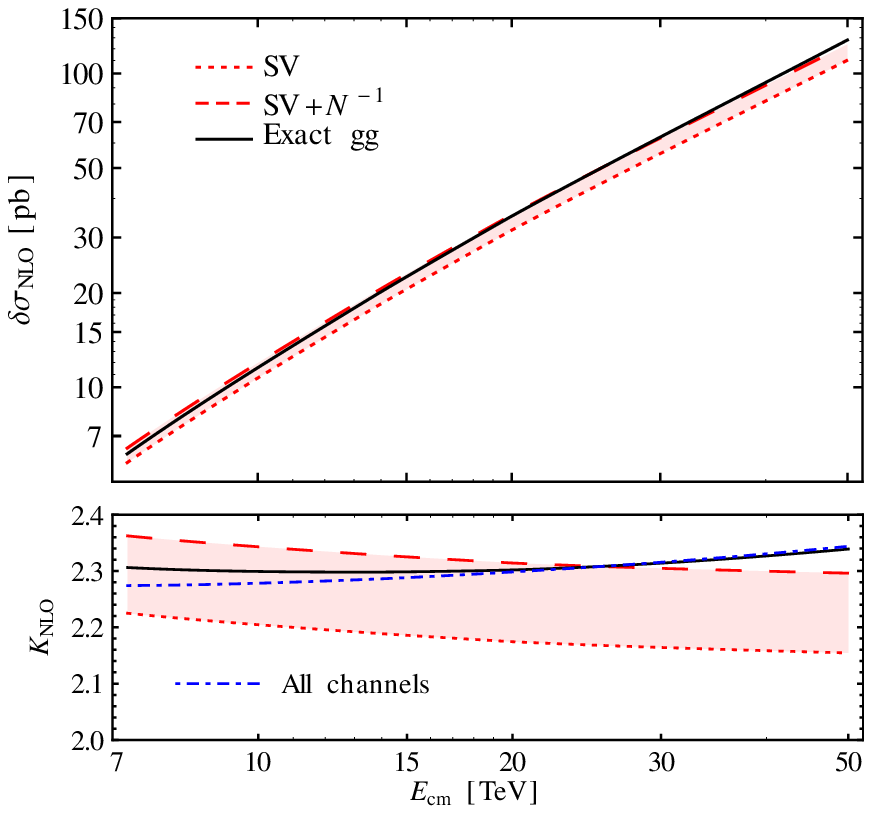}
\\[2mm]
  \includegraphics[width=10.7cm]{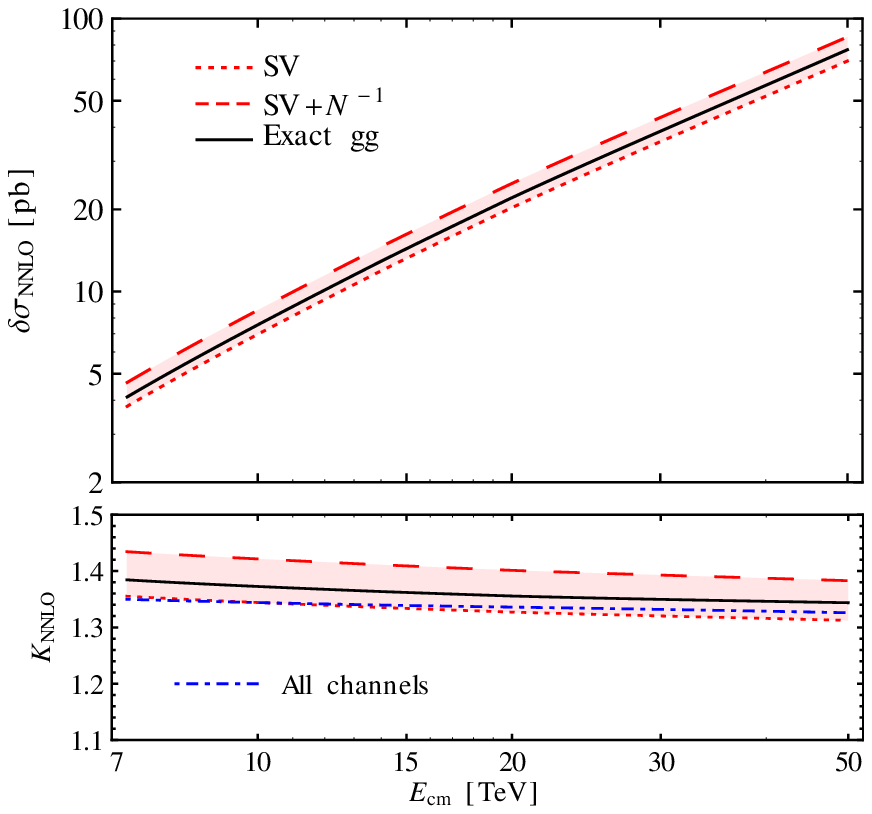}
\end{center}
\vspace{-0.6cm}
\caption{\label{previousXS} \small
 The NLO (top) and NNLO (bottom) gluon-gluon contributions to the Higgs 
 production cross section as a function of the collider energy for the exact  
 coefficient function (solid), the SV approximation (dotted) and the SV terms 
 plus the $N^{\,-1}$ contributions (dashed).
 The lower panels show the corresponding $K$-factors, including the impact of 
 the other partonic subprocesses (dash-dotted).
 All~curves have been calculated using the central NNLO $\as$ and parton 
 densities of Ref.~\cite{Martin:2009iq}.}
\end{figure}

The consistency of the bands in Fig.~\ref{previous} with the exact results at
$N \gsim 2$ does not guarantee the same for the hadronic cross sections at high
collider energies $E_{\rm cm}$. 
Hence we show in Fig.~\ref{previousXS} the NLO and NNLO gluon-gluon 
contributions to the cross section (\ref{eq:had}) for a wide range of 
$E_{\rm cm}$. 
Here and below we have used the exact top-quark mass dependence at LO instead 
of the constant $\widetilde{\sigma}_0^{}$ in Eq.~(\ref{eq:sigma0}) but for now,
as in Table~\ref{table}, the NNLO MSTW \cite{Martin:2009iq} parton set and its 
$\as$ value irrespective of the order of the calculation.
Also displayed in the figure are the results for the corresponding 
`$K$-factors' at NLO and NNLO, 
$ K_{\rm N^kLO} \,=\, \sigma_{\rm N^kLO}/\sigma_{\rm N^{k-1}LO\,}$, 
where we show the rather small (but not negligible) negative effect of the 
quark-gluon and quark-(anti)$\,$quark contributions as well.

We observe that the exact results, for both gluon-gluon fusion and all 
channels, are consistent with the band defined by the SV and SV$\,+\, N^{\,-1}$ 
approximations for $E_{\rm cm} \lsim 20 \mbox{ TeV}$ at NLO (the deviation 
from it remains small even at higher energies) and at all energies considered 
at NNLO, where the approximations are applicable down to somewhat lower values 
of $N$ as shown in Fig~\ref{previous}.
The effect of the non-SV gluon-gluon terms is largely compensated by the other 
channels at NNLO.

\begin{figure}[tb]
\begin{center}
  \includegraphics[width=10.7cm]{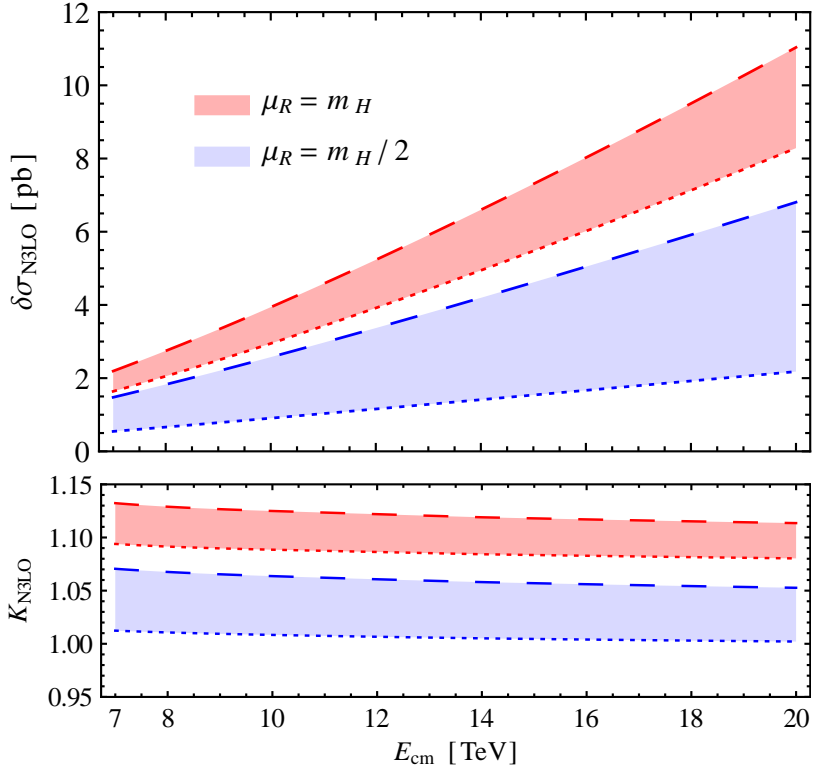}
\end{center}
\vspace{-0.6cm}
\caption{\label{n3loXS}\small
 The N$^3$LO contribution to the Higgs production cross section as a function 
 of the collider energy for the SV approximation (dotted) and the SV terms plus
 the ${\cal O}(N^{-1})$ contributions in Eq.~(\ref{eq:c3est}) (dashed) 
 at~$\muR=\mH$ (upper curves) and $\muR=0.5\,\mH$ (lower curves) for the NNLO 
 gluon distribution of Ref.~\cite{Martin:2009iq} at~$\muF=\mH$.
 The lower panel shows the ratio of these N$^3$LO predictions to the complete
 NNLO result. }
\vspace{-0.3cm}
\end{figure}

In view of these results, we can reliably employ our approximations of 
$c_{gg}^{(3)}(N)$ to predict the size of the N$^3$LO corrections for
$E_{cm}\lsim 20~\rm{TeV}$, as shown in Fig.~\ref{n3loXS}. Here all partonic 
channels are included up to NNLO, while at N$^3$LO we consider only the 
gluon-gluon process. 
The N$^3$LO scale dependence of $\as$ \cite{beta3a,beta3b} has been used with 
$\as(\MZs) = 0.1165$ in the latter case with, since there are no PDF 
parametrizations at this order yet, the NNLO PDFs of Ref.~\cite{Martin:2009iq} 
at the scale $\mHs$.

Under these conditions, the N$^3$LO cross sections are larger at $\muR=\mH$ 
that their NNLO counterparts by
{$11.3\,\% \pm 1.9\,\%$} at $E_{\rm cm} = 7~\rm{TeV}$ and
{$9.7\,\% \pm 1.7\,\%$} at $E_{\rm cm} = 20~\rm{TeV}$. 
At $\muR=\mH/2$, which is closer to the point of minimal sensitivity and 
provides a scale choice that closely reproduces the effect of threshold 
resummation \cite{deFlorian:2012yg}, the corrections are substantially smaller 
with {$4.1\,\% \pm 2.9\,\%$} and {$2.7\,\% \pm 2.5\,\%$}, 
respectively, at 7 TeV and 20 TeV.
Hence the size and present uncertainty of the N$^3$LO corrections is only 
weakly dependent of the collider energy in this range, the latter amounting 
to about 2-3\% at these natural values of $\muR$.

\begin{figure}[p]
\vspace*{-1mm}
\begin{center}
  \includegraphics[width=10.4cm]{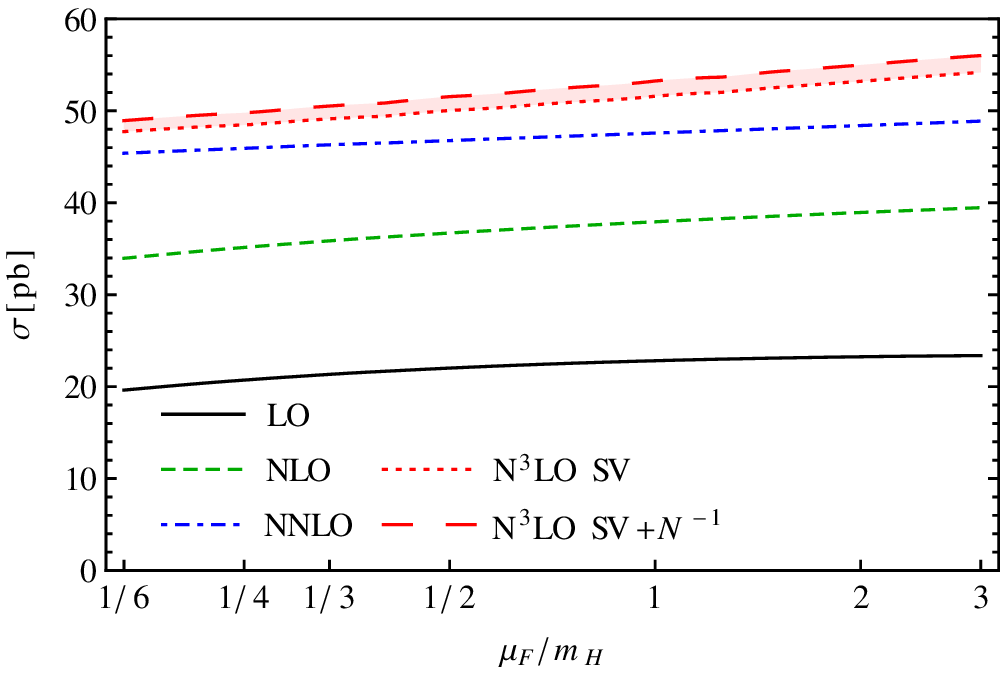}
\\
  \includegraphics[width=10.4cm]{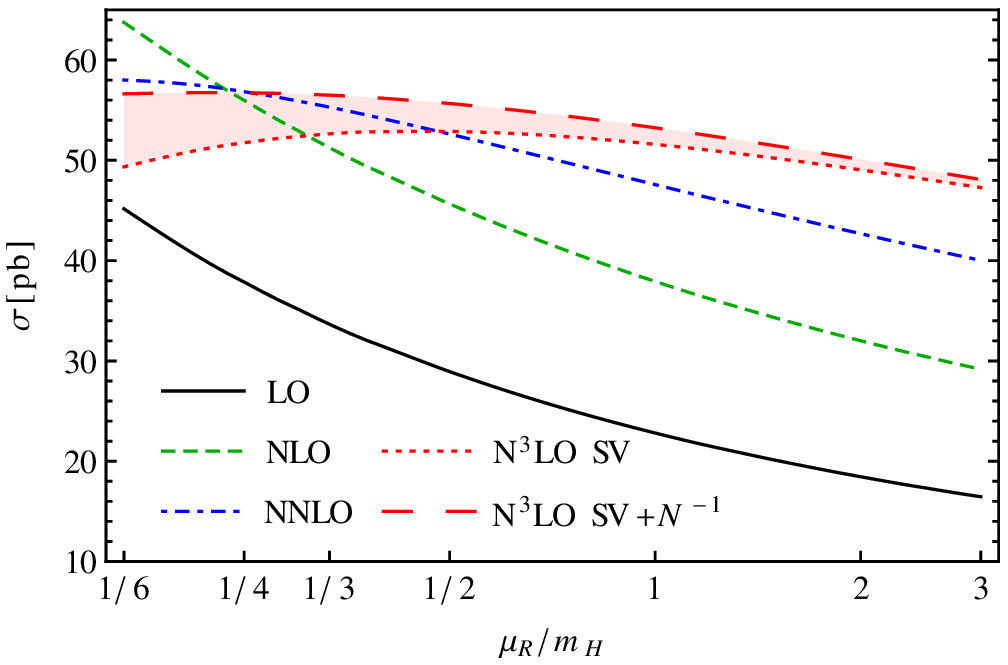}
\\
  \includegraphics[width=10.4cm]{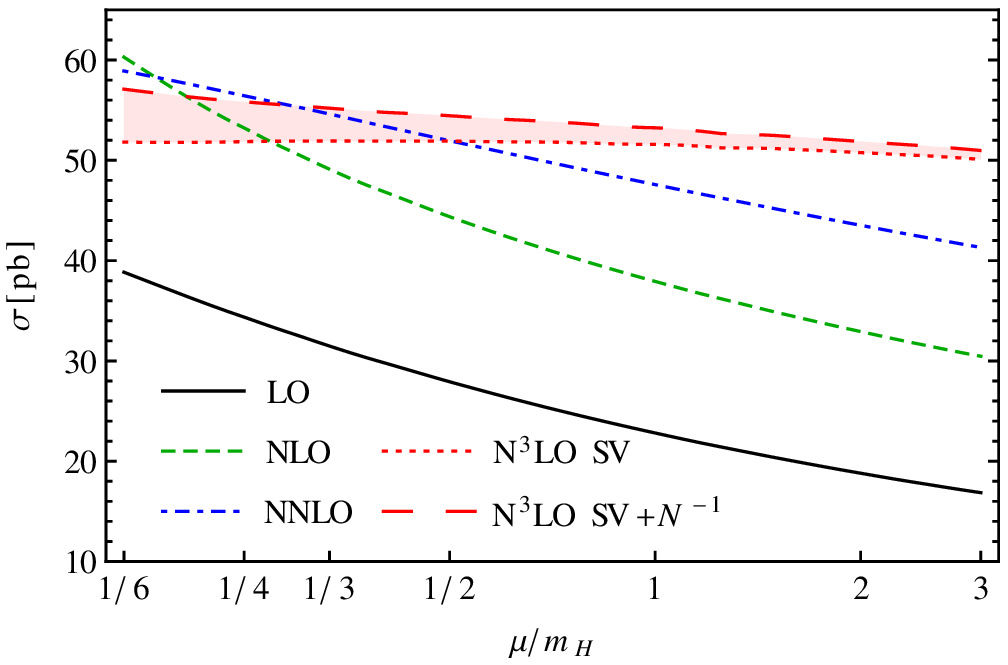}
\end{center}
\vspace{-0.7cm}
\caption{\label{scales} \small
 The dependence of the Higgs production cross section on the factorization 
 scale $\muF$ for $\muR = \mH$ (top), the renormalization scale $\muR$ for
 $\muF = \mH$ (middle), and on $\mu\,\equiv\,\muF=\muR$ (bottom) at 
 $E_{\rm cm}\,=\,14~\rm{TeV}$. 
 Our N$^3$LO band defined by the SV and SV$\,+\,N^{\,-1}$ approximations for
 the coefficient function $c_{gg}^{(3)}(N)$ is compared to the LO, NLO and 
 NNLO results for the respective PDFs and $\as$ values of 
 Ref.~\cite{Martin:2009iq}.} 
\end{figure}

Fig.~\ref{scales} displays the dependence of the total cross section on the 
renormalization and factorization scales $\muR$ and $\muF$ for the successive
perturbative orders, now consistently calculated using (where possible) the
corresponding values and evolution of $\as$ and the PDFs, at 14 TeV.
As shown in the upper plot, the variation with $\muF$ for fixed $\muR$ is small
already at LO, despite the PDFs changing considerably over the wide range of 
scales used in the plots.
The dependence on $\muF$ is, in fact, larger at N$^3$LO than at NNLO; this is 
due to the (presently unavoidable) use of the NNLO gluon distributions also at
this order and the omission of the quark-gluon and quark-(anti)$\,$quark 
channels.

No such caveats apply to the dependence on $\muR$ for fixed $\muF$ which at
N$^3$LO requires `only' the four-loop beta function \cite{beta3a,beta3b} but 
not the so far unknown fourth-order splitting functions. 
Using the interval $0.25\,\mH \leq \muR \leq 2\,\mH$, the cross section ranges 
from {32 to 56~pb} at NLO, from {42.5 to 57~pb} at NNLO, and from 
{49.5 to 54.5~pb} for the center of our N$^3$LO uncertainty band. 
The~respective lower numbers change to {38, 47.5 and 52.5~pb} 
if a more conventional variation by a factor of 2 is used about the apparently 
preferred scale $\mH/2$.
These results indicate an uncertainty due to the truncation of the perturbation
series at N$^3$LO of slightly less than $\pm\, 5\,\%$.

Finally, in the bottom plot in Fig.~\ref{scales}, $\muF$ and $\muR$ are varied
together relative to $\mH$. 
The resulting scale dependence of the cross sections at LO, NLO and NNLO is 
similar to, but slightly smaller than, those just discussed.
The further improvement at N$^3$LO can not be trusted quantitatively, as~the
falling trend towards large scales with $\muR$ is combined with the partly
spurious (see above) increase with $\muF$ shown in the upper plot. 
Hence it is best, at least for the time being, to use the results for a fixed 
$\muF$ for a conservative error estimate.

While often unavoidable, error estimates using scale variations are, of course, 
not particularly reliable; they summarize rather what is known than what will 
be added by yet unknown higher orders, and (width of) the scale range 
considered is somewhat arbitrary. 
A useful alternative is to estimate, where possible, the size to the next order
in the perturbative expansion at a standard scale (for other approaches 
see~\cite{Cacciari:2011ze,David:2013gaa}).
In~the case at hand this is 
possible, since the size of the complete SV contribution at N$^4$LO has been
determined in terms of two parameters that can be estimated, see 
Eq.~(\ref{eq:c4gN}).
In line with the discussion at the end of Section 2, we use $D_{g,4} = 0$ and 
$\kappa_4^{}\, g_{0,4}^{} = 130$ for a `large' estimate of the N$^4$LO 
gluon-gluon coefficient function, and $\kappa_4^{}\, D_{g,4} = 55$, i.e., twice
the Pad\'e approximant in Eq.~(\ref{eq:d4pade}) and $g_{0,4}^{} = 0$ for a 
`small' estimate 
(recall that $\kappa_4 = 1/25000$ effectively converts the fourth-order
quantities to an expansion in $\as$). 

In principle, the N$^4$LO cross section in the SV limit also involves the 
$\alpha_{\rm s}^{\,5}$ contribution to the constant $C(\muRs)$ in Eq.~%
(\ref{eq:sigma0}) which, in fact, is known except for the $n_f$-dependent part 
of the five-loop beta-function of QCD \cite{Schroder:2005hy,Chetyrkin:2005ia}.
However, as obvious from the last row of Table~\ref{table}, this contribution 
can be safely neglected in the present context.

\begin{figure}[tb]
\begin{center}
  \includegraphics[width=10.7cm]{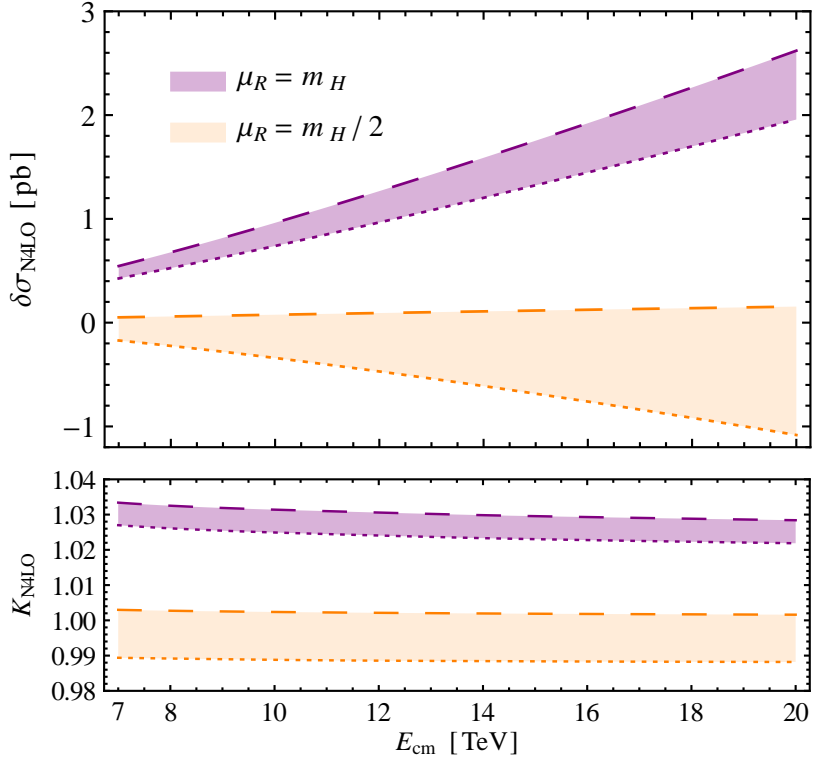}
\end{center}
\vspace{-0.5cm}
\caption{\label{n4loXS}\small 
 As Fig.~\ref{n3loXS}, but for the N$^4$LO corrections as obtained from the 
 `large' and `small' SV estimates of the coefficient function $c_{gg}^{(4)}(N)$ 
 discussed in the text. In the lower panel the N$^4$LO results are shown 
 relative to the corresponding N$^3$LO cross sections in the SV approximation.}
\end{figure}

The resulting estimates for the N$^4$LO correction are shown in 
Fig.~\ref{n4loXS} in the same manner as the N$^3$LO contributions in 
Fig.~\ref{n3loXS}. Also here the relative size to the corrections depends 
weakly on the colliders energy between 7 TeV and 20 TeV, with about
{$3.0\,\%$ to $2.5\,\%$} at \mbox{$\muR = \mH$} and
{$-0.4\,\%$ to $-0.5\,\%$} at $\muR = \mH/2$.
At $E_{\rm cm} = 14 \mbox{ TeV}$ the N$^4$LO SV terms change the respective
N$^3$LO cross sections by about {{1.5 pb}} and 
{{-0.5 pb}}. 
Even if these results were to considerably underestimate the true N$^4$LO 
correction, the latter would still amount to less than {5\%}. 
%
Note that the bands here and in Fig.~\ref{n3loXS} above have to be added 
(upper panels), or are shown relative to (lower panels), the different
lower-order results at the two scales. Hence the difference between the 
bands for $\muR = \mH$ and $\muR = \mH/2$ does not indicate the overall
scale uncertainty of the N$^3$LO and N$^4$LO predictions.

In view of these and the above results, a combined perturbation-series 
uncertainty of about $\pm\, 5\,\%$ can be assigned to our present N$^3$LO
cross section, which takes into account the approximate character of 
$c_{gg}^{(3)}(N)$, the omission of the N$^3$LO quark-gluon and quark-(anti)%
$\,$quark contributions and the truncation of the expansion at this point. 
Calculating all higher-order contributions in the heavy-top approximation but 
normalizing with the full lowest-order result, this leads to a total cross 
section of {$54.3 \pm 2.7$ pb} 
at 14 TeV for the NNLO PDFs of Ref.~\cite{Martin:2009iq} -- which should under-
or overestimate the corresponding N$^3$LO gluon-gluon luminosity by less than
{1\%} -- and $\as(\MZs) = 0.1165$, where the central value 
refers the choice $\muR = \mH/2$ and $\muF = \mH$. As all our results, the
%
above cross section does not include either electroweak corrections or 
bottom-mass effects.

Our present result for the N$^3$LO corrections in the SV approximation is 
larger, by about a factor of two at $\muR = \mH$, than that given ten years 
ago in Ref.~\cite{MV2005}. 
This is due to the recently calculated coefficient of $\delta \zt$ 
\cite{Anastasiou:2014vaa}, which turns out to be almost twice as large as
anticipated for the uncertainty estimate in Ref.~\cite{MV2005}, and the 
different input parameters, most notably a larger value of $\as(\MZs)$.
Our results including the $N^{\,-1} \ln^{\,k\!}N$ term in Eq.~(\ref{eq:c3est})
can be compared to Refs.~\cite{Ball:2013bra,Bonvini:2014jma}, where an 
approximate N$^3$LO prediction has been constructed, based on the large and
small-$N$ behavior of the partonic cross section (for which the latter has a
small effect at LHC energies). As mentioned above, their $N^{\,-1} \ln^{\,k}N$ 
terms due not agree with our result except for the obvious coefficient of 
$N^{-1}\ln^{\,5} N$. Nevertheless, the central prediction of 
Refs.~\cite{Ball:2013bra,Bonvini:2014jma} for the N$^3$LO cross section is
rather comparable to our result.

Finally, with the perturbative QCD corrections to the coefficient function of 
the dominant hard scattering process well under control, the largest remaining 
uncertainties in predictions of the physical cross section originate from the 
input parameters for $\as$ and the PDFs, cf.~Eq.~(\ref{eq:had}). 
For instance, use of the ABM12 value of $\as$ and PDFs~\cite{Alekhin:2013nda}, 
which were tuned to LHC data, leads to central values for the cross section 
which are significantly lower, by some 11-14\% (depending on the collider 
energy), than those reported, e.g., in Table~\ref{table} and Fig.~\ref{scales}, 
see~Ref.~\cite{Alekhin:2013nda}.
This is due to a smaller value of $\as(\MZs)$ and a smaller gluon distribution
in the relevant $z$-range for the ABM12 parametrization as compared to MSTW 
\cite{Martin:2009iq}; the origin of these differences has been understood 
\cite{Alekhin:2011ey,Alekhin:2012ig}.
Very recently, also the NNPDF collaboration has reported new and slightly 
lower values of the Higgs cross section for the NNPDF$\:$3.0 parton set~\cite{
Rojo@ichep:2014} also tuned to LHC data.

%
\section{Summary and outlook}

For almost ten years rigorous results for the total Higgs-production cross
section in the heavy top-quark limit have been confined to the exact NNLO
coefficient functions 
\cite{Harlander:2002wh,Anastasiou:2002yz,Ravindran:2003um}
plus the N$^3$LL soft-gluon resummation 
\cite{MV2005,Laenen:2005uz,Idilbi:2005ni}
which fixes the highest six threshold logarithms at all higher orders.
Earlier this year an N$^3$LO diagram calculation has been completed in the 
soft$\,+\,$virtual limit \cite{Anastasiou:2014vaa}, adding the coefficient of 
$\delta \zt$ to those of the $[\zt^{-1} \ln^{\,k} \zt]_+$ terms with 
$0 \le k \le 5$.
 
Progress has also been made in the past years on resumming sub-dominant
large-$z$ logarithms, $\zt^a \ln^{\,k} \zt$ with $a \geq 0$, via physical
evolution kernels \cite{MV2009c,SMVV} or the structure of unfactorized cross
sections in dimensional regularization \cite{AV2010,ASV}; the latter
has been used recently to derive the leading large-$z$ logs for the
quark-gluon contribution to Higgs production to all orders~\cite{LPAV2014}.

Here we have considered the dominant gluon-gluon channel and extended the
calculations of Ref.~\cite{MV2009c} to Higgs-boson production. Based on the
results of Refs.~\cite{Harlander:2002wh,Anastasiou:2002yz,Ravindran:2003um}
we have thus derived the leading sub-SV contributions, $\ln^{\,k} \zt$ with 
$k = 5,\,4,\,3$, the first two completely (unsurprisingly verifying the 
conjecture of Ref.~\cite{Kramer:1996iq} for the leading logarithm) and the 
third up to a constant of minor numerical relevance.
The corresponding results for $a \geq 1$ can only be derived for the 
\mbox{non-$\,C_F$} terms at this point, consequently only the coefficient of 
the leading logarithms is complete. 
These results, included in the Appendix together with their fourth-order 
counterparts, can provide a non-trivial check on a future complete N$^3$LO 
calculation.

Switching to Mellin moments for phenomenological considerations, a comparison 
of the pattern of the coefficients at NLO, NNLO and N$^3$LO with those for 
Higgs-exchange DIS, where the coefficient function is fully known to N$^3$LO 
\cite{SMVV}, allows to give well-motivated estimates for the remaining 
$N^{\,-1} \ln^{\,2,\,1,\,0}N$ third-order contributions to $c_{gg}^{(3)}(N)$.
It turns out that both the corresponding coefficient functions at
$N \gsim 2$ as well as the NLO and NNLO contributions to the cross sections for
LHC energies are contained in a band spanned by the respective SV and 
SV$+{\cal O}(N^{-1})$ approximations. 
Assuming the same situation at the third order, we have been able to improve 
upon previous estimates \cite{MV2005,Ball:2013bra} of the size and remaining 
uncertainty of the N$^3$LO correction.

We have studied the dependence of these approximate N$^3$LO results on the 
renormalization and factorization scales, as well as the size of the N$^4$LO 
corrections in the SV approximation. 
We conclude that the remaining perturbation-series uncertainty amounts to no 
more than~$\pm 5\%$, which includes the effects of approximate character of 
$c_{gg}^{(3)}(N)$, the omission of the N$^3$LO quark-gluon and quark-(anti)%
$\,$quark contributions and the truncation of the series. 
Using the central NNLO PDFs of Ref.~\cite{Martin:2009iq} at $\muF=\mH$ and the 
N$^3$LO strong coupling with $\as(\MZs) = 0.1165$ leads to an increase by
{$(10 \pm 2)\%$} at $\muR = \mH$ and {$(3 \pm 2.5)\%$} at $\muR = \mH/2$, 
which appears to be the preferred central scale, over the corresponding NNLO
cross sections at a collider energy of 14 TeV.

\pagebreak

The perturbative expansion of the hard scattering cross section is, therefore,
now quite well under control, rendering the uncertainties of the PDFs and 
$\as$ an at least as important source of uncertainties for LHC predictions.
Given the progress on the perturbative QCD corrections reported in 
Ref.~\cite{Anastasiou:2014vaa} and here,
together with new global fits of PDFs to LHC data, it appears that the cross 
section values~\cite{Heinemeyer:2013tqa} recommended for use in the ongoing
and upcoming ATLAS and CMS Higgs analyses
require revision, for Run2 of the LHC, to include the latest theory 
developments and improvements on the evaluation of the parton distributions 
and the value of $\as$.

 
\appendix
\section{Large-$N$ expansions at N$^3$LO and N$^4$LO}
\renewcommand{\theequation}{{\rm{A}}.\arabic{equation}}
\setcounter{equation}{0}

Here we present the general expressions corresponding to Eqs.~(\ref{eq:c3gN}) 
and (\ref{eq:c4gN}).
For compactness the results are written in terms of $\ln \,\Ntil \,=\, 
\ln \,N + \GE$.  The $N^{\,0}$ coefficients at N$^3$LO read 
{\small
\bea
\label{eq:cn3sge}
  c_{gg}^{\,(3)} \Big|_{\,\ln^{\,6}\Ntil}  &\!\! = \! &
          {256 \over 3}\: \* \cat 
\nn \:\: , \\[2mm]
  c_{gg}^{\,(3)} \Big|_{\,\ln^{\,5}\Ntil} &\!\! = \! &
         { 1408 \over 9 }\: \* \cat  
       - { 256 \over 9 }\: \* \cas \* \nf  
\nn \:\: , \\[2mm]
  c_{gg}^{\,(3)} \Big|_{\,\ln^{\,4}\Ntil} &\!\! = \! &
         \cat  \*  \left[
            { 14800 \over 27 }
          + 384\, \* \z2
          \right]
       - { 2624 \over 27 }\: \* \cas \* \nf 
       + { 64 \over 27 }\: \* \ca \* \nfs 
\nn \:\: , \\[2mm]
  c_{gg}^{\,(3)} \Big|_{\,\ln^{\,3}\Ntil} &\!\! = \! &
         \cat  \*  \left[
            { 67264 \over 81 }
          - 448\, \* \z3
          + { 704 \over 3 }\: \* \z2
          \right]
       - \cas \* \nf  \*  \left[
            { 14624 \over 81 }
          + { 128 \over 3 }\: \* \z2
          \right]
       - { 32 \over 3 }\: \* \ca \* \cf \* \nf  
       + { 640 \over 81 }\: \* \ca \* \nfs  
\nn \:\: , \\[2mm]
  c_{gg}^{\,(3)} \Big|_{\,\ln^{\,2}\Ntil} &\!\! = \! &
        \cat  \*  \left[
            { 122276 \over 81 }
          + { 15008 \over 9 }\: \* \z2
          - { 5632 \over 9 }\: \* \z3
          + { 2752 \over 5 }\: \* \zss
          \right]
       - \cas \* \nf  \*  \left[
            { 33688 \over 81 }
          + { 2240 \over 9 }\: \* \z2
          + { 704 \over 9 }\: \* \z3
          \right]
\nn \\ & & \mbox{}
       - \ca \* \cf \* \nf  \*  \Big[
            252
          - 192\, \* \z3
          \Big]
       + { 800 \over 81 }\: \* \ca \* \nfs  
\nn \:\: , \\[2mm] 
 c_{gg}^{\,(3)} \Big|_{\,\ln^{\,1}\Ntil} &\!\! = \! &
         \cat  \*  \left[
            { 594058 \over 729 }
          + { 64784 \over 81 }\: \* \z2
          - { 24656 \over 27 }\: \* \z3
          - { 176 \over 5 }\: \* \zss
          - { 2336 \over 3 }\: \* \z2\, \* \z3
          + 384\, \* \z5
          \right]
\nn \\[1mm] & & \mbox{\hspn}
       + \cas \* \nf  \*  \left[
          - { 125252 \over 729 }
          - { 9104 \over 81 }\: \* \z2
          + { 1808 \over 27 }\: \* \z3
          - { 32 \over 5 }\: \* \zss
          \right]
       - \ca \* \cf \* \nf  \*  \left[
            { 3422 \over 27 }
          - { 608 \over 9 }\: \* \z3
          - { 64 \over 5 }\: \* \zss
          \right]
\nn \\[1mm] & & \mbox{\hspn}
       + \ca \* \nfs  \*  \left[
            { 3712 \over 729 }
          + { 64 \over 9 }\: \* \z3
          \right]
\nn \:\: , \\[2mm]
 c_{gg}^{\,(3)} \Big|_{\,\ln^{\,0}\Ntil} &\!\! = \! &
         \cat  \*  \left[
            { 215131 \over 81 }
          + { 186880 \over 81 }\: \* \z2
          - { 130828 \over 81 }\: \* \z3
          + { 119692 \over 135 }\: \* \zss
          - { 2024 \over 3 }\: \* \z2\, \* \z3
          + { 3476 \over 9 }\: \* \z5
\right. \nn \\[1mm] & & \left. \mbox{\hspp} 
          + { 3872 \over 15 }\: \* \zst
          + 96\, \* \zts
          \right]
       \:\: + \:\: \ca \* \nfs  \*  \left[
            { 2515 \over 27 }
          - { 1328 \over 81 }\: \* \z2
          + { 3344 \over 81 }\: \* \z3
          - { 224 \over 15 }\: \* \zss
          \right]
\nn \\[1mm] & & \mbox{\hspn} 
       + \cas \* \nf  \*  \left[
          - { 98059 \over 81 }
          - { 38168 \over 81 }\: \* \z2
          + { 296 \over 81 }\: \* \z3
          - { 4696 \over 135 }\: \* \zss
          - { 784 \over 3 }\: \* \z2\, \* \z3
          + { 808 \over 9 }\: \* \z5
          \right]
\nn \\[1mm] & & \mbox{\hspn} 
       + \ca \* \cf \* \nf  \*  \left[
          - { 63991 \over 81 }
          - { 3404 \over 9 }\: \* \z2
          + { 1184 \over 3 }\: \* \z3
          + { 176 \over 45 }\: \* \zss
          + 384\, \* \z2\, \* \z3
          + 160\, \* \z5
          \right]
\nn \\[1mm] & & \mbox{\hspn} 
       + \cfs \* \nf  \*  \left[
            { 608 \over 9 }
          + { 592 \over 3 }\: \* \z3
          - 320\, \* \z5
          \right]
       + \cf \* \nfs  \*  \left[
            { 8962 \over 81 }
          - { 184 \over 9 }\: \* \z2
          - { 224 \over 3 }\: \* \z3
          - { 32 \over 45}\: \* \zss
          \right]
\:\: .
\eea
}
Except for the $\ln^{\,0}\Ntil$ part, these results have been presented before 
in a different notation, e.g., in Appendix E of Ref.~\cite{Catani:2003zt}. 
Our new $N^{\,-1}$ terms read, with one unknown coefficient $\xiH$
of Eq.~(\ref{eq:xiH})
{\small
\bea
\label{eq:cn3corr}
  c_{gg}^{\,(3)} \Big|_{\,N^{\,-1}\ln^{\,5}\Ntil}  &\!\! = \! &
          256\, \* \cat
\nn \:\: , \\[1mm]
  c_{gg}^{\,(3)} \Big|_{\,N^{\,-1}\ln^{\,4}\Ntil} &\!\! = \! &
         { 7552 \over 9 }\: \* \cat
       - { 640 \over 9}\: \* \cas \* \nf
\:\: , \\[1mm]
  c_{gg}^{\,(3)} \Big|_{\,N^{\,-1}\ln^{\,3}\Ntil} &\!\! = \! &
         \cat  \*  \left[
            { 29312 \over 27 }
          + { 11 \over 9 }\: \* \xiH
          + 768\, \* \z2
          \right]
       + \cas \* \nf  \*  \left[
          - { 4960 \over 27 }
          - { 2 \over 9 }\: \* \xiH
          \right]
       + { 128 \over 27 }\: \* \ca \* \nfs  
\nn \:\: .
\eea
}
The corresponding N$^4$LO results are given by
{\small
\bea
\label{eq:cn4sge}
  c_{gg}^{\,(4)} \Big|_{\,\ln^{\,8}\Ntil}  &\!\! = \! &
          {512 \over 3}\: \* \caf
\nn \:\: , \\[3mm]
  c_{gg}^{\,(4)} \Big|_{\,\ln^{\,7}\Ntil} &\!\! = \! &
         { 5632 \over 9 }\: \* \caf  
       - { 1024 \over 9 }\: \* \cat \* \nf  
\nn \:\: , \\[3mm]
  c_{gg}^{\,(4)} \Big|_{\,\ln^{\,6}\Ntil} &\!\! = \! &
         \caf  \*  \left[
            { 216320 \over 81 }
          + { 2560 \over 3 }\: \* \z2
          \right] 
       - { 45568 \over 81 }\: \* \cat \* \nf  
       + { 2048 \over 81 }\: \* \cas \* \nfs 
\nn \:\: , \\[3mm]
  c_{gg}^{\,(4)} \Big|_{\,\ln^{\,5}\Ntil} &\!\! = \! &
        \caf  \*  \left[
            { 838112 \over 135 }
          + { 14080 \over 9 }\: \* \z2
          - 1792\, \* \z3
          \right]
       - \cat \* \nf  \*  \left[
            { 26048 \over 15 }
          + { 2560 \over 9 }\: \* \z2
          \right]
       - { 256 \over 3 }\: \* \cas \* \cf \* \nf  
\nn \\[1mm] & & \mbox{\hspn} 
       + { 17024 \over 135 }\: \* \cas \* \nfs  
       - { 256 \over 135 }\: \* \ca \* \nft  
\nn \:\: , \\[3mm]
  c_{gg}^{\,(4)} \Big|_{\,\ln^{\,4}\Ntil} &\!\! = \! &
         \caf  \*  \left[
            { 3450592 \over 243 }
          - { 45056 \over 9 }\: \* \z3
          + { 250912 \over 27 }\: \* \z2
          + { 7936 \over 5 }\: \* \zss
          \right]
\nn \\[1mm] & & \mbox{\hspn}
       + \cat \* \nf  \*  \left[
          - { 1084592 \over 243 }
          - { 1024 \over 9 }\: \* \z3
          - { 41600 \over 27 }\: \* \z2
          \right]
       + \cas \* \cf \* \nf  \*  \left[
          - { 12592 \over 9 }
          + 1024\, \* \z3
          \right]
\nn \\[1mm] & & \mbox{\hspn}
       + \cas \* \nfs  \*  \left[
            { 77152 \over 243 }
          + { 640 \over 27 }\: \* \z2
          \right]
       + { 160 \over 9 }\: \* \ca \* \cf \* \nfs
       - { 640 \over 81 }\: \* \ca \* \nft
\nn \:\: , \\[3mm]
  c_{gg}^{\,(4)} \Big|_{\,\ln^{\,3}\Ntil} &\!\! = \! &
         \caf  \*  \left[
            { 13631360 \over 729 }
          + { 923968 \over 81 }\: \* \z2
          - { 1125184 \over 81 }\: \* \z3
          + { 7040 \over 9 }\: \* \zss
          - { 16000 \over 3 }\: \* \z2\, \* \z3
          + 3072\, \* \z5
          \right]
\nn \\[1mm] & & \mbox{\hspn}
       + \cat \* \nf  \*  \left[
          - { 4591096 \over 729 }
          - { 219904 \over 81 }\: \* \z2
          + { 116096 \over 81 }\: \* \z3
          - { 11008 \over 45 }\: \* \zss
          \right]
       + { 16 \over 3 }\: \* \ca \* \cfs \* \nf
\nn \\[1mm] & & \mbox{\hspn}
       + \cas \* \cf \* \nf  \*  \left[
          - 2208
          - 128\, \* \z2  
          + { 3968 \over 3 }\: \* \z3
          + { 512 \over 5 }\: \* \zss
          \right] 
       + \ca \* \cf \* \nfs  \*  \left[
            { 5600 \over 27 }
          - { 1280 \over 9 }\: \* \z3
          \right]
\nn \\[1mm] & & \mbox{\hspn}
       + \cas \* \nfs  \*  \left[
            { 436760 \over 729 }
          + { 1280 \over 9 }\: \* \z2
          + { 7424 \over 81 }\: \* \z3
          \right]
       - { 3200 \over 243 }\: \* \ca \* \nft
\nn \:\: , \\[3mm]
  c_{gg}^{\,(4)} \Big|_{\,\ln^{\,2}\Ntil} &\!\! = \! &
         \caf  \*  \left[
            { 28356478 \over 729 }
          + { 2800672 \over 81 }\: \* \z2
          - { 799888 \over 27 }\: \* \z3
          + { 873104 \over 135 }\: \* \zss
          - { 82720 \over 9 }\: \* \z2\, \* \z3
\right. \nn \\[1mm] & & \left. \mbox{\hspp}
          + { 65824 \over 9 }\: \* \z5
          + { 25792 \over 15 }\: \* \zst
          + 2336\, \* \zts
          \right]
       \:\: + \:\: \ca \* \cfs \* \nf  \*  \left[
            { 4864 \over 9 }
          - 2560\, \* \z5
          + { 4736 \over 3 }\: \* \z3
          \right]
\nn \\[1mm] & & \mbox{\hspn}
       + \cat \* \nf  \*  \left[
          - { 12176488 \over 729 }
          - { 661136 \over 81 }\: \* \z2
          + 3152\, \* \z3
          - { 32768 \over 135 }\: \* \zss
          - { 19520 \over 9 }\: \* \z2\, \* \z3
          - { 448 \over 9 }\: \* \z5
          \right]
\nn \\[1mm] & & \mbox{\hspn}
       + \cas \* \cf \* \nf  \*  \left[
          - { 751982 \over 81 }
          - { 34576 \over 9 }\: \* \z2
          + { 15232 \over 3 }\: \* \z3
          + { 7744 \over 45 }\: \* \zss
          + 3840\, \* \z2\, \* \z3
          + 1280\, \* \z5
          \right]
\nn \\[1mm] & & \mbox{\hspn}
       + \cas \* \nfs  \*  \left[
            { 1072784 \over 729 }
          + { 11680 \over 81 }\: \* \z2
          + { 9760 \over 27 }\: \* \z3
          - { 320 \over 3 }\: \* \zss
          \right]
       + \ca \* \nft  \*  \left[
          - { 7424 \over 729 }
          - { 128 \over 9 }\: \* \z3
          \right]
\nn \\[1mm] & & \mbox{\hspn}
       + \ca \* \cf \* \nfs  \*  \left[
            { 110996 \over 81 }
          - { 2624 \over 3 }\: \* \z3
          - { 1472 \over 9 }\: \* \z2
          - { 1408 \over 45 }\: \* \zss
          \right]
       \:\: + \:\: 2\, \* \Ag4 
\nn \:\: , \\[3mm]
  c_{gg}^{\,(4)} \Big|_{\,\ln^{\,1}\Ntil} &\!\! = \! &
         \caf  \*  \left[
            { 50096 \over 9 }
          + { 29565664 \over 729 }\: \* \z2
          + { 2426936 \over 243 }\: \* \z3
          - { 1592288 \over 405 }\: \* \zss
          - { 876608 \over 27 }\: \* \z2\, \* \z3
\right. \nn \\[1mm] & & \left. \mbox{\hspp}
          + { 4928 \over 5 }\: \* \zst
          + { 17248 \over 9 }\: \* \zts
          - { 9824 \over 3 }\: \* \zss \* \z3
          + 6144\, \* \z2\, \* \z5
          \right]
       + 16\, \* \z2\, \* \ca \* \cfs \* \nf  
\nn \\[1mm] & & \mbox{\hspn}
       + \cat \* \nf  \*  \left[
          - { 191776 \over 81 }
          - { 10159592 \over 729 }\: \* \z2
          - { 1819648 \over 243 }\: \* \z3
          + { 820928 \over 405 }\: \* \zss
          + { 127616 \over 27 }\: \* \z2\, \* \z3
\right. \nn \\[1mm] & & \left. \mbox{\hspp}
          + { 4928 \over 9 }\: \* \zts
          - 384\, \* \zst
          \right]
       \:\: + \:\: \ca \* \cf \* \nfs  \*  \left[
            { 15008 \over 81 }
          + 384\, \* \z2
          + { 256 \over 27 }\: \* \z3
          - 256\, \* \z2\, \* \z3
          \right]
\nn \\[1mm] & & \mbox{\hspn}
       + \cas \* \cf \* \nf  \*  \left[
          - { 108272 \over 81 }
          - { 116096 \over 27 }\: \* \z2
          + { 38504 \over 27 }\: \* \z3
          + 128\, \* \zss
          + { 22400 \over 9 }\: \* \z2\, \* \z3
\right. \nn \\[1mm] & & \left. \mbox{\hspp}
          + { 1024 \over 5 }\: \* \zst
          - 896\, \* \zts
          \right]
       \:\: + \:\: \ca \* \nft  \*  \left[
          - { 3200 \over 81 }\: \* \z2
          - { 5120 \over 81 }\: \* \z3
          + { 256 \over 45 }\: \* \zss
          \right]
\nn \\[1mm] & & \mbox{\hspn}
       + \cas \* \nfs  \*  \left[
            { 17920 \over 81 }
          + { 1019464 \over 729 }\: \* \z2
          + { 349184 \over 243 }\: \* \z3
          - { 10624 \over 45 }\: \* \zss
          \right]
       \:\: - \:\: \Dg4
\eea
}
with the yet unknown fourth-order quantities $A_{g,4}$ and $D_{g,4}$, and
{\small
\bea
\label{eq:cn4corr}
  c_{gg}^{\,(4)} \Big|_{\,N^{\,-1}\ln^{\,7}\Ntil}  &\!\! = \! &
         { 2048 \over 3 }\: \* \caf
\nn \:\: , \\[1mm]
  c_{gg}^{\,(4)} \Big|_{\,N^{\,-1}\ln^{\,6}\Ntil} &\!\! = \! &
         { 35840 \over 9 }\: \* \caf  
       - { 3584 \over 9 }\: \* \cat \* \nf  
\:\: , \\[1mm]
  c_{gg}^{\,(4)} \Big|_{\,N^{\,-1}\ln^{\,5}\Ntil} &\!\! = \! &
         \caf  \*  \left[
            { 244736 \over 27 }
          + 2560\, \* \z2
          + { 88 \over 9 }\: \* \xiH
          \right]
       - \cat \* \nf  \*  \left[
            { 49792 \over 27 }
          + { 16 \over 9 }\: \* \xiH
          \right]
       + { 2048 \over 27 }\: \* \cas \* \nfs 
\nn \:\: .
\eea
}
The corresponding $N^{\,0}$ contributions for the DY process can now be
written down at the same accuracy due to the determination of the coefficient
of $\delta \zt$ at N$^3$LO in Refs.~\cite{Ahmed:2014cla,Catani:2014uta}.
The DY counterparts of Eqs.~(\ref{eq:cn3corr}) and (\ref{eq:cn4corr}) have
been determined in Ref.~\cite{MV2009c};
the leading $\ln^{\,2k-1}\zt$ terms at $k$-loops of those agree with the result
of Ref.~\cite{Laenen:2008ux}.


\section{z-space results beyond (1$\,$--$\,$z)$^0$ for large z}
\renewcommand{\theequation}{{\rm{B}}.\arabic{equation}}
\setcounter{equation}{0}

For non-singlet quantities such as the dominant quark-antiquark annihilation
contribution to the total cross section for Drell-Yan lepton-pair production,
$pp/p\bar{p} \to l^+l^-+X$, the physical kernel is single-log enhanced at all 
orders in the expansion about $z\!=\!1$ \cite{MV2009c}. 
This is also true for the $C_A^{\,k} \, n_{\! f}^{\,\ell}$ contributions to
Higgs production via gluon-gluon fusion in the heavy-top limit, viz
\bea
\label{eq:Kggx}
 K_{gg}^{\,(1)}(z)
  & \!=\! &
  \ln \zt \,\pgg(z) \* \left[
      - 16\,\ca \beta_0 - 32\,\cas\, \H(0) \right]
      \;+\; {\cal O} \! \left( \:\!\ln^{\,0} \!\zt \:\!\right) \:\: ,
\nn \\[2mm]
 K_{gg}^{\,(2)}(z)
  & \!=\! &
      \ln^{\,2} \zt \,\pgg(z)\* \left[ \:
      32\,\ca \beta_0^{\,2} + 112\,\cas\,\beta_0\, \H(0)
      + 128\,\cat\: \Hh(0,0) \right]
      \;+\; {\cal O} \! \left( \:\!\ln \zt \:\!\right)
\nn \:\: ,
\\[2mm]
 K_{gg}^{\,(3)}(z)
  & \!=\! &
      \ln^{\,3} \zt \,\pgg(z)\* \left[
      - 64\,\ca \beta_0^{\,3}
      - \xi_H^{\,(3)}\, \cas\,\beta_0^{\,2}\, \H(0)
      - \eta_H^{\,(3)}\, \cat\,\beta_0\, \Hh(0,0)  
      - \xi_P^{\,(3)}\, \caf\, \Hh(0,0,0) \right]
\nn \\
 & & \mbox{}
      +\; {\cal O} \! \left( \:\!\ln^{\,2} \!\zt \:\!\right)
\eea
at $\,\muR = \mH$ with $ \H(0) \,=\, \ln z\,$, 
$\Hh(0,0) \,=\, 1/2\: \ln^{\,2} \! z\,$, 
$\Hhh(0,0,0) \,=\, 1/6\: \ln^{\,3} \! z$ \cite{Remiddi:1999ew} and
\[
  \pgg(z) \:\: =\:\: \zt_+^{\,-1} - 2 + z^{\,-1} + z - z^{\,2}
\:\: .
\]
The first two lines of Eq.~(\ref{eq:Kggx}) are a direct consequence of
Refs.~\cite{Dawson:1990zj,Djouadi:1991tka} and~\cite
{Anastasiou:2002yz,Harlander:2002wh,Ravindran:2003um}; their numerical
coefficients are the same as for the Drell-Yan case in Eq.~(3.27) of Ref.~\cite
{MV2009c}, which is based on the results of Refs.~\cite
{Hamberg:1991np,Harlander:2002wh}, up to a factor of two due to the different 
normalizations of $p_{gg}$ here and~$p_{qq}$ in Ref.~\cite{MV2009c}. 
The N$^3$LO generalization based on the results for DIS, where the 
corresponding coefficient functions are known \cite{SMVV,MVV6,MVV10},
involves two presently unknown parameters of the third-order coefficient
function, $\xi_H^{\,(3)}$ already encountered above and $\eta_H^{\,(3)}$
relevant at $\zt^{k \geq 1}$,
and one unknown coefficient of the four-loop splitting function 
$P_{gg}^{\,(3)}$ which is not relevant here.

Eq.~(\ref{eq:Kggx}) together with Eqs.~(\ref{eq:physkern}) and 
(\ref{eq:ctilde}) above yields the $\muF \,=\/\muR \,=\, \mH$ results
\bea
\label{cH3allz}
\lefteqn{  
  4^{\,-3}\, c_{gg}^{(3)}(z)\Big|_{\,\cf \,=\, 0} \: = \:\: 
  \left(
          \lnzt5 \: \* 8\, \* \cat
  \, - \, \lnzt4 \: \* 10/3\: \* \cas \* \bz
  \, + \, \lnzt3 \: \* 1/3\: \* \ca \* \bn2
  \right) \* \pgg(z) 
} \quad \nn \\ & & \mbox{\hspn}
  + \lnzt4 \: \* \cat\: \* 
  \Big\{
    - 27\, \* \H(0)\, \* \pgg(z) 
    - 32\, \* \H(0)\, \* (1+z) 
    + 59\, \* (1-z)
    - 187/3\: \* \left (z^{\,-1\!} - z^{\,2} \right)
  \! \Big\}
\nn \\[1mm] & & \mbox{\hspn}
  + \lnzt3 \, \* \cat \*
  \Big\{
    \left[
      16/3 
    - 56\, \* \z2 
    + \left( 170/3 + \etaH\!/96 \right) \* \Hh(0,0) 
    \right] \* \pgg(z)
\nn \\[1mm] & & \mbox{}
    + \left[
      4\, \* \Hh(0,0) 
    - 8\, \* \THh(-1,0)
    \right] \* \pgg(-z)
    - \left(
      119
    - 407/3\: \* z^{\,-1}
    - 205\, \* z
    + 605/3\: \* z^{\,2}
    \right) \* \H(0)
\nn \\[1mm] & & \mbox{}
    + \left(
      76
    + 140\, \* z
    \right) \* \Hh(0,0) 
    - 128\, \* (1+z)\, \* \THh(1,0)
    - 721/3
    + 2875/12\: \* z
    + 2314/9\: \* \left (z^{\,-1\!} - z^2 \right)
  \! \Big\}
\quad \nn \\[1mm] & & \mbox{\hspn}
  + \lnzt3 \: \* \cas \* \bz\: \*
  \Big\{ \!
    \left(
      20/3\: \* (1+\H(0))
    + \xiH\!/192\: \* \H(0)
    \right) \* \pgg(z)
    + 10\, \* (1+z)\, \* \H(0)
    - 67/3 
\nn \\[1mm] & & \mbox{}
    + 271/12\: \* z
    + 193/9\: \* \left (z^{\,-1\!} - z^{\,2} \right)
  \! \Big\}
  \;\; + \;\; {\cal O}\left(\ln^{\,2} \zt \right)
\eea
and
\bea
\label{cH4allz}
\lefteqn{
  4^{\,-4}\, c_{gg}^{(4)}(z)\Big|_{\,\cf \,=\, 0} \: = \:\:
  \left(
    \lnzt7 \: \* 16/3\: \* \caf
  - \lnzt6 \: \* 14/3\: \* \cat \* \bz
  + \lnzt5 \: \* 4/3\: \* \cas \* \bn2
  \right) \* \pgg(z)
} \quad  \nn \\ & & \mbox{\hspn}
  + \lnzt6 \: \* \caf\: \*
  \Big\{
    - 77/3\: \* \H(0)\, \* \pgg(z)
    - 32\, \* (1+z)\, \* \H(0)
    + 166/3\: \* (1-z)
    - 550/9\: \* \left (z^{\,-1\!} - z^{\,2} \right)
  \! \Big\}
\quad \nn \\[1mm] & & \mbox{\hspn}
  + \lnzt5 \: \* \caf\: \*
  \Big\{
    \left[
      8 
    - 92\, \* \z2 
    + \left( 244/3 + \etaH\!/96 \right) \* \Hh(0,0)
    \right] \* \pgg(z)
\nn \\[1mm] & & \mbox{}
    + \left[
      4\, \* \Hh(0,0)
    - 8\, \* \THh(-1,0)
    \right] \* \pgg(-z)
    - \left(
      156 
    - 220\, \* z^{\,-1}
    - 306\, \* z
    + 286\, \* z^{\,2}
    \right) \* \H(0)
\nn \\[1mm] & & \mbox{}
    + \left(
      104
    + 232\, \* z
    \right) \* \Hh(0,0)
    - 192\, \* (1+z)\, \* \THh(1,0)
    - 1265/3
    + 5051/12\: \* z
    + 3818/9 \: \* \left (z^{\,-1\!} - z^{\,2} \right)
  \! \Big\}
\nn \\[1mm] & & \mbox{\hspn}
  + \lnzt5 \: \* \cat \* \bz\: \*
  \Big\{ \!
      \left[ 10 + \left( 91/6 + \xiH\!/96 \right) \* \H(0)
      \right] \* \pgg(z)
    + 70/3\: \* (1+z)\, \* \H(0)
    - 265/6
\nn \\[1mm] & & \mbox{}
    + 533/12\: \* z
    + 93/2 \* \left (z^{\,-1\!} - z^{\,2} \right)
  \! \Big\}
  \:\: + \:\: {\cal O}\left(\ln^{\,4} \zt \right)
\; .
\eea
Here we have again suppressed the argument $z$ of the harmonic polylogarithms
for which we use a partly modified basis in terms of functions that have Taylor 
expansions about $z=1$ with rational coefficients \cite{MV2009c}
including
\bea
  \THh(1,0)(z) \; & = & \; \Hh(1,0)(z) \: + \: \z2
  \quad\:\: = \; - \ln z \, \ln \zt - \mbox{Li}_2(z) + \z2
\:\: , \nn \\[1mm]
  \THh(-1,0)(z) & = & \Hh(-1,0)(z) + \z2/2
  \; = \:\: \ln z \, \ln (1\!+\!z) + \mbox{Li}_2(-z) + \z2/2
\:\: .  \nn
\eea
Similar to their NNLO analogues 
\cite{Anastasiou:2002yz,Harlander:2002wh,Ravindran:2003um} 
and the NNLO and N$^3$LO coefficient function for Higgs-exchange DIS 
\cite{SMVV}, the complete coefficient functions corresponding to 
Eqs.~(\ref{cH3allz}) and (\ref{cH4allz}) will include additional 
$C_F$-terns contributing from $\zt^1$ beyond the leading logarithms.
 
The corresponding results for the non-singlet quark-antiquark annihilation 
contribution to the Drell-Yan process are given by$\,$%
\footnote{The $\ln^{\,4\!} \zt$ and $\ln^{\,5\!} \zt$ contributions to 
$\,c_{\rm DY}^{(3)\,\rm ns}(z)\,$ have been presented before in Eq.~(6.29) of 
Ref.~\cite{MV2009c} where, unfortunately, all coefficients are too small 
by a factor 3/4.}
\bea
\label{cDY3allz}
\lefteqn{
  4^{\,-3}\, c_{\,\rm DY}^{\,(3)\,\rm ns}(z) \: = \:\:
  \left(
        \lnzt5 \: \* 4\, \* \cft
  \,-\, \lnzt4 \: \* 5/3\: \* \cfs \* \bz
  \,+\, \lnzt3 \: \* 1/6\: \* \cf \* \bn2
  \right) \* \pqq(z)
}
\quad \nn \\[1mm] & & \mbox{\hspn}
  + \lnzt4 \: \* \cft\: \*
  \Big\{
    - 27/2\: \* \H(0)\, \* \pqq(z)
    + 4\, \* (1+z)\, \* \H(0)
    - 8\, \* (1-z)
  \! \Big\}
\nn \\[1mm] & & \mbox{\hspn}
  + \lnzt3 \, \* \cft \*
  \Big\{
    \left[
    - 16 
    - 24\, \* \z2 
    - 3\, \* \H(0) 
    - \THh(1,0)
    + \left( 79/3 + \etaD/192 \right) \* \Hh(0,0)
    \right] \* \pqq(z)
\nn \\[1mm] & & \mbox{}
    + \left(
      17/2
    - 73/2\: \* z
    \right) \* \H(0)
    - 27/2\: \* (1+z)\, \* \Hh(0,0)
    + 14\, \* (1+z)\, \* \THh(1,0)
    + 8 
    - 17/2\: \* z
  \! \Big\}
\nn \\[1mm] & & \mbox{\hspn}
  + \lnzt3 \: \* \cfs \* \bz\: \*
  \Big\{ \!
    \left[
      10/3 
    + \left( 13/3 + \xiD/384 \right) \* \H(0)
    \right] \* \pqq(z)
    - (1+z)\, \* \H(0)
    + 4\,\* (1-z)
  \! \Big\}
\nn \\[1mm] & & \mbox{\hspn}
  + \lnzt3 \: \* \cfs \* \ca\: \*
  \Big\{ \!
    \left(
      8/3 
    - 4\,\* \z2 
    + \THh(1,0) 
    + 2\,\* \Hh(0,0) 
    \right) \* \pqq(z)
    + (1+z)\, \* ( \THh(1,0) + 2\,\* \H(0) )
\nn \\[1mm] & & \mbox{}
    + 6 - 11/2\, \* z
  \Big\}
  \:\: + \:\: {\cal O}\left(\ln^{\,2} \zt \right)
\eea
and
\bea
\label{cDY4allz}
\lefteqn{
  4^{\,-4}\, c_{\,\rm DY}^{\,(4)\,\rm ns}(z) \: = \:\:
  \left(
        \lnzt7 \: \* 8/3\: \* \cff
  \,-\, \lnzt6 \: \* 7/3\: \* \cft \* \bz
  \,+\, \lnzt5 \: \* 2/3\: \* \cfs \* \bn2
  \right) \* \pqq(z)
} \nn \\[1mm] & & \mbox{\hspn}
  + \lnzt6 \: \* \cff\: \*
  \Big\{
    - 77/6\: \* \H(0)\, \* \pqq(z)
    + 4\, \* (1+z)\, \* \H(0)
    - 8\, \* (1-z)
  \! \Big\}
%
\nn \\[1mm] & & \mbox{\hspn}
  + \lnzt5 \, \* \cff \*
  \Big\{
    \left[
    - 16
    - 40\, \* \z2
    - 3\, \* \H(0)
    - \THh(1,0)
    + \left( 116/3 + \etaD/192 \right) \* \Hh(0,0)
    \right] \* \pqq(z)
\nn \\[1mm] & & \mbox{}
    + \left(
      16
    - 52\, \* z
    \right) \* \H(0)
    - 21\: \* (1+z)\, \* \Hh(0,0)
    + 22\, \* (1+z)\, \* \THh(1,0)
    + 16
    - 33/2\: \* z
  \! \Big\}
\nn \\[1mm] & & \mbox{\hspn}
  + \lnzt5 \: \* \cft \* \bz\: \*
  \Big\{ \!
    \left[
      5
    + \left( 103/12 + \xiD/192 \right) \* \H(0)
    \right] \* \pqq(z)
    - 8/3\: \* (1+z)\, \* \H(0)
    + 22/3\: \* (1-z)
  \! \Big\}
\nn \\[1mm] & & \mbox{\hspn}
  + \lnzt5 \: \* \cft \* \ca\: \*
  \Big\{ \!
    \left(
      4
    - 6\,\* \z2
    + \THh(1,0)
    + 2\,\* \Hh(0,0)
    \right) \* \pqq(z)
    + (1+z)\, \* ( \THh(1,0) + 2\,\* \H(0) )
\nn \\[1mm] & & \mbox{}
    + 6 - 11/2\, \* z
  \Big\}
  \:\: + \:\: {\cal O}\left(\ln^{\,4} \zt \right)
\eea
with 
$$
  \pqq(z) \;=\; 2\, \zt_+^{\,-1} - 1 - z \:\: .
$$
The $\lnzt3$ term in Eq.~(\ref{cDY3allz}) and the $\lnzt5$ contribution in 
Eq.~(\ref{cDY4allz}) include the unknown third-order coefficients 
$\xi_{\rm DY}^{(3)}$ and $\eta_{\rm DY}^{(3)}$ which we definitely expect to be
equal to their counterparts for Higgs-boson production in Eqs.~(\ref{eq:Kggx}) 
-- (\ref{cH4allz}).
Hence an extension of either Refs.~\cite
{Anastasiou:2002yz,Harlander:2002wh,Ravindran:2003um} or Refs.~\cite
{Hamberg:1991np,Harlander:2002wh} to N$^3$LO will fix also the third-highest
power of $\ln \zt$ at N$^4$LO and all higher orders for both processes. 


\subsection*{Acknowledgments}

\vspace*{-2mm}
\noindent
This work has been supported by the European Union through contract
PITN-GA-2010-264564 ({\it LHCPhenoNet$\,$})
and by the UK {\it Science \& Technology Facilities Council}$\,$ (STFC) under
grant~number ST/G00062X/1.

\vspace*{-3mm}
{\small
\setlength{\baselineskip}{0.36cm}

}

\end{document}